\documentclass[
 reprint,
 amsmath,
 amssymb,
 aip,
floatfix,
longbibliography
]{revtex4-2}

\usepackage{graphicx}
\usepackage{dcolumn}
\usepackage{bm}
\usepackage[x11names]{xcolor}
\usepackage[percent]{overpic}

\usepackage{pict2e}

\usepackage{siunitx}
\DeclareSIUnit\molar{\text{M}}

\usepackage[normalem]{ulem}

\begin{document}


\newcommand{\conc}[1]{c_{\text{#1}}}
\newcommand{\Kd}[1]{K_{\mathrm{D}}^{\mathrm{#1}}}
\newcommand{\Ka}[1]{K_{\mathrm{a}}^{\mathrm{#1}}}
\newcommand{\DF}{\gamma}
\newcommand{\koff}[1]{k_{\mathrm{off}}^{\mathrm{#1}}}
\newcommand{\kon}[1]{k_{\mathrm{on}}^{\mathrm{#1}}}
\newcommand{\kb}{k_{\mathrm{B}}}
\newcommand{\kT}{k_{\mathrm{B}}T}
\newcommand{\cc}[1]{c_{\text{#1}}}
\newcommand{\ts}{\textsuperscript}
\newcommand{\diff}[2]{\frac{\mathrm{d} #1}{\mathrm{d} #2}}
\newcommand{\uM}{\si{\micro\molar}}


\title{Hybrid Dynamical Simulation Reveals Apparent Stiffening of Flexible Protein Lattices Driving Membrane Bending}

\author{Samuel L. Foley}
  \email{sfoley13@jhu.edu}
  \affiliation{T. C. Jenkins Department of Biophysics, Johns Hopkins University, Baltimore, Maryland, USA}
\author{Margaret E. Johnson}%
  \email{margaret.johnson@jhu.edu}
  \affiliation{T. C. Jenkins Department of Biophysics, Johns Hopkins University, Baltimore, Maryland, USA
}

\date{\today}

\begin{abstract}
Membrane-deforming protein lattices play a central role in essential and pathogenic remodeling processes, including clathrin-mediated endocytosis and viral budding. Simulating these systems at biologically relevant length and time scales requires mesoscale approaches that preserve structural detail while avoiding the computational cost of atomistic resolution. Here, we present a hybrid simulation framework that couples a particle-based flexible protein lattice to a continuum membrane model, enabling systematic investigation of how lattice geometry and rigidity influence dynamic membrane remodeling. We validate the coupled model by comparing simulation results with theoretical predictions for membranes under increasing tension. Using buckling-based deformations of pre-assembled clathrin lattices, we quantify the lattice flexural rigidity and establish a direct relationship between the force constants in the coarse-grained energy function and the emergent lattice mechanics. We then compare this flexural rigidity, $\kappa_{l,\mathrm{flex}}$, to an effective rigidity, $\kappa_{l,\mathrm{eff}}$, commonly used in continuum descriptions of sphere-forming protein assemblies. Although $\kappa_{l,\mathrm{flex}}$ is set solely by the energy function, $\kappa_{l,\mathrm{eff}}$ depends on lattice size and connectivity, with the two measures converging only for weakly-connected lattices. As a result, with lattice growth, the effective rigidity relevant for spherical bud formation increases and the relaxation timescale decreases. This size-dependent stiffening highlights the importance of structural details in interpreting lattice mechanics and cautions against assuming a single constant stiffness. We further demonstrate the generality of the method by applying it to pre-assembled viral lattices generated with reaction-diffusion simulations. Overall, this work provides a validated framework for simulating how deformable protein assemblies of diverse geometry couple to membrane dynamics and remodeling.
\end{abstract}

\maketitle

\section{Introduction}

Lipid bilayer membranes are essential semipermeable barriers present in all living organisms.
Far from being static structures, membranes constantly undergo bending, reshaping, and remodeling in response to both active cellular processes and passive environmental forces \cite{bassereau20182018,kosmalska2015physical}.
Many of these remodeling events are mediated by cytosolic proteins that bind directly to the membrane or to membrane-associated adaptor proteins \cite{fu2023modeling,nishimura2018membrane,pfitzner2021principles,traub2003sorting,vassilopoulos2024clathrin}.
In processes such as viral budding and endocytosis, including the well-studied clathrin-mediated endocytosis (CME), semi-rigid protein assemblies dynamically form at the membrane and act as scaffolds that promote curvature generation and shape evolution \cite{cocucci2012first,willy2021novo}.
Because these processes involve coupled protein mechanics, membrane elasticity, and lattice growth, simulation methods that can resolve these physical interactions in time provide critical tools for predicting the determinants of membrane reshaping.
A primary goal here is to introduce a general hybrid mesoscale simulation method that captures these coupled dynamics for preassembled lattices of varying sizes, connectivity, and protein subunit types.
A key outcome of this study is that we map the parameters of the elastic energy function to the emergent rigidity of the protein lattice in two ways, including the apparent observed rigidity of inducing a spherical membrane cap parameterized by $\kappa_{l,\mathrm{eff}}$.
With this approach, we demonstrate how the connectivity and size of the lattice can strongly impact their apparent stiffness, $\kappa_{l,\mathrm{eff}}$ for driving membrane budding.

While computational studies of protein-driven membrane remodeling offer a powerful means of testing how the tunable properties of protein lattices and membranes combine to control shape change, simulating these processes at biologically relevant length and time scales remains challenging. Atomistic and near-atomistic methods are too expensive for large membrane systems with slow remodeling dynamics \cite{fu2023modeling,deserno2009mesoscopic}.
This is true for membranes of even moderate size: for example, a vesicle with radius 50 nm already contains more than $10^5$ lipids, beyond the scale of all-atom simulations aimed at membrane remodeling.
One way to extend accessible time and length scales is to coarse-grain (CG) the membrane into larger particles interacting through effective pair potentials \cite{beiter2024making}.
Depending on the level of coarse-graining, these simulations have been used to study multi-protein curvature induction \cite{liebl2025membrane,beiter2026subdomains}, and the dynamics of filament-driven cell division\cite{vanhille2024self,munoz2025tutorial}, viral budding \cite{ruiz2015simulations,yuan2010one}, and membrane-mediated aggregation \cite{reynwar2007aggregation}.
Even so, these approaches can remain computationally costly, and the coarse-grained interactions must still be carefully tuned to reproduce experimentally relevant membrane properties such as bending rigidity \cite{fu2023modeling,sorichetti2026beads}.
An alternative and more efficient approach is to replace explicit diffusing lipids with a surface-based description of the membrane, using either a continuum mesh \cite{siggel2022trimem,brakke1992surface,zhu2022mem3dg,feng2006finite,fu2021continuum,badvaram2023physical} or a bead-spring surface \cite{giani2017early,gompper1997network,ding2019shapes,davtyan2017mesoscopic,kumar2022membrane,li2024shaping}.
In the present work, we employ a continuum membrane model that is both computationally efficient and directly parameterized by experimentally measurable membrane properties, including bending rigidity and tension \cite{lin2004brownian}.
By propagating the membrane in Fourier space, the model can also account for hydrodynamic effects efficiently. The key limitation of this Fourier-space Brownian dynamics (FSBD) formulation is that it is valid only in the small-gradient limit of the Helfrich functional \cite{seifert1997configurations,deserno2015fluid}.
We therefore focus on the early stages of vesicle formation, before strongly curved $\Omega$-shaped deformations develop.
Although FSBD has previously been used with proteins represented either as a continuum or as individual particles, here we couple it to flexible, particle-based elastic lattices.

A major advantage of representing the proteins with explicit coarse molecular structure is that it allows the relationship between microscopic energy parameters in the CG model and emergent, meso-scale lattice mechanics to be examined directly. In particular, we can ask how lattice size and connectivity influence the apparent rigidity of a stable protein assembly as it drives membrane remodeling. This is important because many continuum descriptions of membrane-deforming protein assemblies assume that the lattice can be represented by a constant effective bending modulus. For a protein assembly modeled as an elastic thin-shell continuum, however, the full elastic energy generally includes bending, Gaussian curvature, and stretching contributions \cite{roos2010physical,lidmar2003virus}.
In simplified continuum treatments, often only the quadratic bending term is retained \cite{frey2024coat,tagiltsev2021nanodissected} or additionally the Gaussian modulus \cite{walani2015endocytic,hassinger2017design}, with a single effective bending rigidity assumed throughout assembly.
Our simulations allow this approximation to be tested directly: by assuming a single fixed coarse-grained continuum rigidity for the protein lattice and measuring how the elastic response changes as the lattice bends, we can extract an apparent rigidity associated with membrane deformation.
We find that this effective rigidity is not constant, but instead increases as the lattice grows and becomes more internally connected. Because self-assembly is stochastic, intermediate structures can be partially connected \cite{qian2023temporal}, asymmetric \cite{ying2026membrane}, or defective \cite{guo2022large}, and our results here show that these structural differences can significantly alter the apparent stiffness inferred from membrane-bending behavior.
These findings are consistent with experimental observations of clathrin lattices transitioning from flat to curved morphologies \cite{sochacki2021structure} and they support prior theoretical suggestions that flat-to-curved transitions during growth requires lattice stiffening when interpreted through the simplified continuum model with a single bending modulus \cite{frey2024coat}.

In this paper, we combine structure-resolved rigid protein assemblies generated by the reaction-diffusion simulator NERDSS (NonEquilibrium Reaction-Diffusion Self-assembly Simulator) \cite{varga2020nerdss}
with flexible coarse-grained mechanics and a continuum membrane model.
In contrast to fully reactive simulations of (dis)assembly produced by NERDSS, we focus here on stable, preassembled lattices to isolate the coupled dynamics of membrane deformation and lattice elasticity.
Rigid protein subunits are connected through harmonic potentials that constrain bond lengths, angles, and dihedrals about preferred structural values, enabling rigid-body protein geometry to be embedded in a deformable lattice \cite{frenkel2025understanding}.
These lattices are then coupled to a continuum membrane propagated using FSBD, and the full system is advanced using HOOMD-blue \cite{anderson2020hoomd}.
We show that the harmonic force constants can be calibrated to reproduce a meso-scale rigidity within the range of experimentally measured rigidity parameters using two distinct active-deformation methods, including a buckling-based analysis \cite{hu2013determining,noguchi2011anisotropic}, and we validate that the coupled system responds as expected to applied membrane tension. We quantify the timescales for the coupled membrane-lattice systems to reach their equilibrium state given an initial flat geometry. We show that the lattice curvature equilibrates within microseconds, with faster relaxation for stiffer and more highly-connected lattices. 
Finally, we demonstrate that the same approach can be applied to other assemblies generated by NERDSS, using an HIV-1 immature Gag lattice as a proof of principle \cite{qian2023temporal}.
This work provides a computational framework for studying how stable, deformable protein assemblies of diverse geometry interact with membranes during remodeling.
It also establishes a quantitative connection between coarse-grained interaction parameters and emergent lattice mechanics, while showing that structural details such as connectivity and lattice size can strongly influence the apparent rigidity inferred from membrane deformation.
In doing so, the model provides a route toward fully reactive simulations of dynamic protein-driven membrane bending.

\section{Methods}

\begin{figure*}
    \centering
    \begin{overpic}[width=\linewidth]{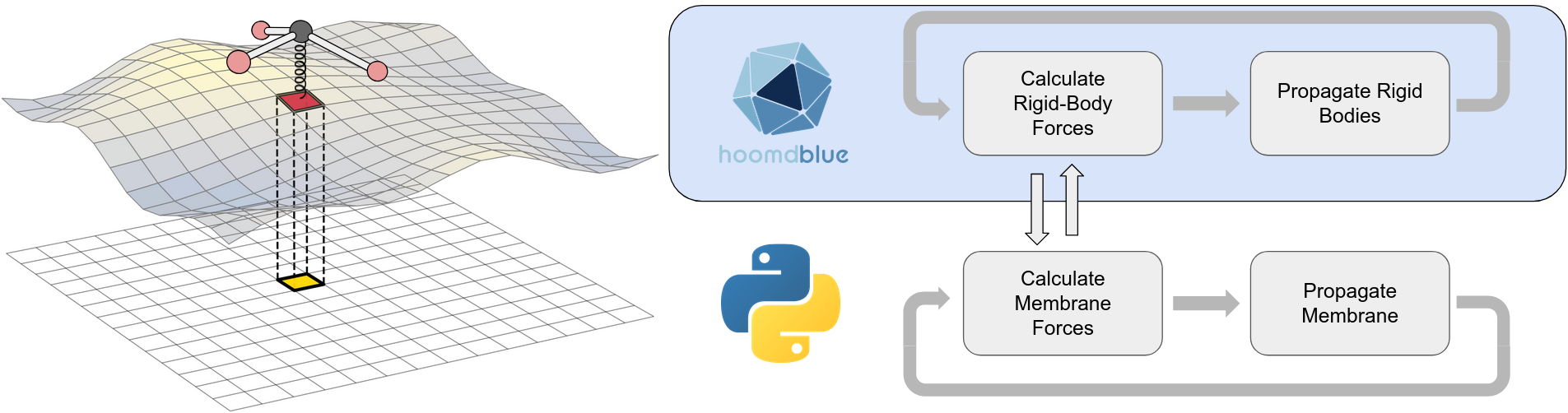}
        \put(2,24){\Large \textbf{a}}
        \put(38,24){\Large \textbf{b}}
    \end{overpic}
    \\
    \begin{overpic}[width=0.5\linewidth]{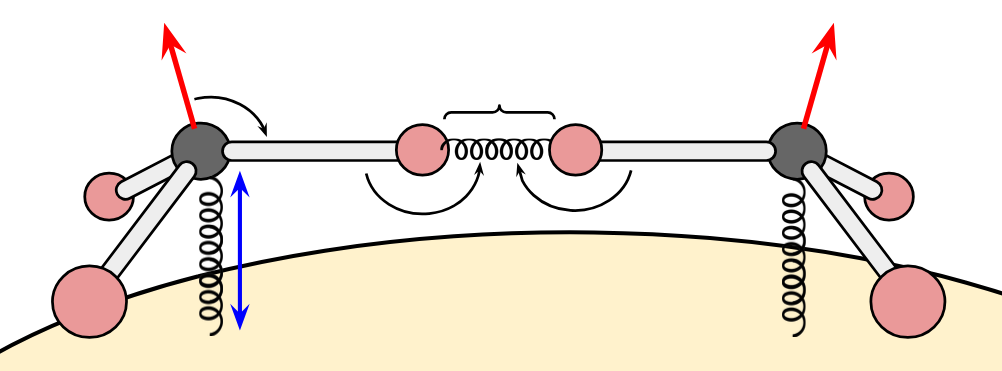}
        \put(4,32){\Large \textbf{c}}
        \put(25,13){\textcolor{blue}{$\Delta z$}}
        \put(23,28){$\alpha$}
        \put(49,28){$\sigma$}
        \put(34,15){$\theta_1$}
        \put(63,15.5){$\theta_2$}
    \end{overpic}
    \begin{overpic}[width=0.25\linewidth]{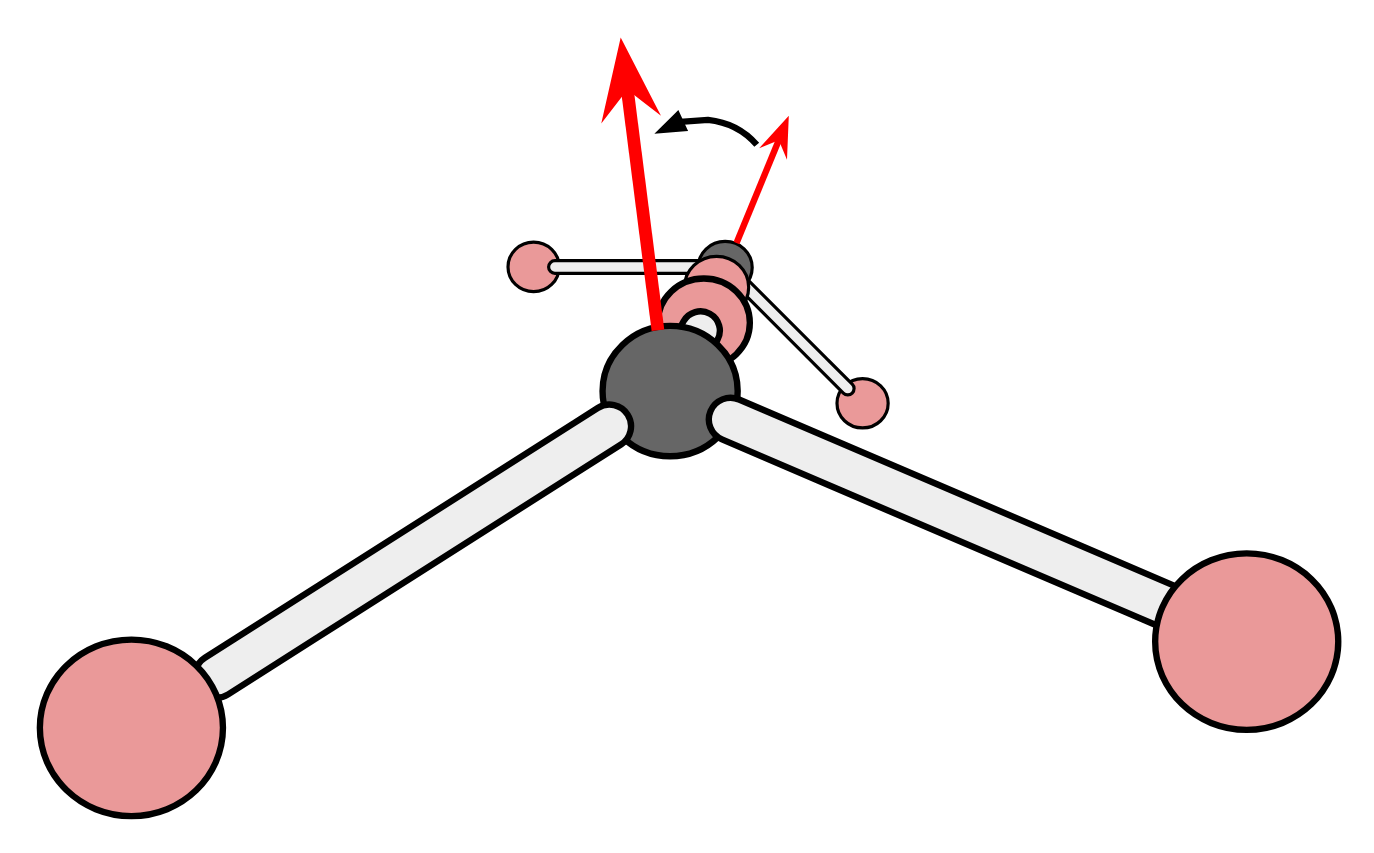}
        \put(50,56){$\omega$}
    \end{overpic}
    \caption{(\textbf{a}) Schematic diagram of protein-membrane coupling, with CG clathrin as an example.
    Individual triskelia are bound to the membrane via harmonic springs in the z-direction.
    This force is applied in the discretized membrane model by contributing a force per unit area to the membrane patch over which the triskelion resides.
    (\textbf{b}) Overview of the hybrid propagation scheme.
    The rigid-body Brownian dynamics are handled by hoomdblue, while the membrane is propagated in Fourier space using our own python code.
    (\textbf{c}) Schematic diagram of CG model parameters for a NERDSS-resolution simulation of clathrin-clathrin association, as well as the membrane-binding degree of freedom $\Delta z$. The right-hand figure shows a rotated view to illustrate the torsion angle $\omega$. }
    \label{fig:overview}
\end{figure*}

\subsection{Flexible Protein Networks}

Our coarse-grained proteins are represented at the same resolution as those of the rigid-body reaction-diffusion software NERDSS \cite{varga2020nerdss} in order to enable inter-operability of these two simulation frameworks.
Here, we recount the coarse-graining setup used for protein-protein binding in NERDSS, and highlight the extra information required to carry out a flexible simulation of those same models coupled to a membrane.

Individual proteins (monomers) are rigid bodies consisting of a center of mass (COM) site along with binding sites to which other proteins may associate.
This is exemplified by the clathrin monomers shown in Fig.~\ref{fig:overview}c, where the COM site is shown as dark grey and the three equally-spaced binding sites $6\mathrm{nm}$ away from the COM are pink.
The pucker angle $\alpha$ encodes a particular monomer curvature preference (conventionally measured from the vertical such that $\alpha=90^\circ$ yields a flat structure).
The geometry of the clathrin monomers is the same as in ref. \cite{guo2022large}, with adaptor binding sites omitted.

In NERDSS, protein-protein association is parametrized by a set of geometric parameters that fully determine the rigid-body structure of the bound pair.
While the individual proteins themselves will remain rigid in our flexible lattices, the bonds and angles between bound pairs will be assigned an energy function.
Here, we will focus on the parameters used to specify the clathrin-clathrin bonds for our clathrin lattices.
These are the site-to-site distance $\sigma$, the two bond angles $\theta_1$ and $\theta_2$, and the torsion angle $\omega$, all illustrated in Fig.~\ref{fig:overview}.
For our clathrin-clathrin bonds, the values are $\sigma=5\,\mathrm{nm}$, $\theta_1 = \theta_2 = 180^\circ$, and $\omega = 0$.
In NERDSS, these parameters yield rigidly defined complexes where subunits ``snap'' into pre-defined orientations following reactive association events. To allow protein complexes defined by these rigid bond parameters to be flexible, we associate to each parameter a harmonic energy
\begin{equation}
    U_x = \frac{1}{2} k_x (x - x_0)^2,
    \label{eqn:harmonic-energy}
\end{equation}
where $x$ can be any of $\sigma$, $\theta$, $\omega$, or $\phi$, with $\phi$ being another dihedral angle required for more general interacting pairs that lack the symmetry and linearity of the clathrin-clathrin bond.
A flexible protein assembly simulation therefore requires a rigidity parameter for each distinct parameter of each unique bond type present in the complex.
For our clathrin coat system, this means we need the three bond rigidity parameters $k_\sigma$, $k_\theta$, and $k_\omega$.
$k_\sigma$ is the spring constant that stabilizes the bond length, $k_\theta$ keeps the leg-to-leg bonds approximately straight, and $k_\omega$ impedes torsional twisting of the triskelia about each bond (see Fig.~\ref{fig:overview}c).
The particular combination of values chosen for these parameters will give rise to the meso-scale elastic rigidity of the clathrin coat.
The ability of a simulation to represent realistic, biologically relevant processes is thus dependent on the calibration of these parameters to experimental data, which we consider in the following sections.
Finally, the protein monomers are coupled to the membrane by harmonic springs, once again with the same form as Eqn.~(\ref{eqn:harmonic-energy}) but with $x = \Delta z$, where $\Delta z$ is the vertical distance between the protein's membrane binding site and the membrane itself.
The membrane force coupling being only in the $z$-direction is a result of our using the small-gradient approximation of the Helfrich energy, which we will discuss in the next section.

The monomeric protein subunits are still treated as rigid bodies within our simulations.
To propagate our rigid-body protein monomers, we use HOOMD-blue's Brownian integrator for both translational and rotational dynamics\cite{anderson2020hoomd,nguyen2011rigid}.
The COM position $\mathbf{x}_i$ of body $i$ follows the equation,
\begin{equation}
    \frac{\mathrm{d}\mathbf{x}_i}{\mathrm{d}t} = \frac{1}{\gamma_i}\left( \mathbf{F}_{\mathrm{net},i} + \mathbf{F}_{\mathrm{R},i} \right),
    \label{eqn:com-brownian}
\end{equation}
where $\mathbf{F}_{\mathrm{net},i}=-\nabla_i U$ is the net force derived from the potential energy, $\mathbf{F}_{\mathrm{R},i}$ is a stochastic force that obeys the appropriate fluctuation-dissipation relation, and $\gamma_i$ is the translational drag coefficient related to the diffusion constant of the species of body $i$ by the Stokes-Einstein relation $D=\kT/\gamma$.
The rotational degrees of freedom for each rigid body, encoded in a rotation quaternion $\mathbf{q}_i$, obey an equivalent equation:
\begin{equation}
    \frac{\mathrm{d}\mathbf{q}_i}{\mathrm{d}t} = \frac{1}{\gamma_{r,i}} \left( \boldsymbol{\tau}_{\mathrm{net},i} + \boldsymbol{\tau}_{\mathrm{R},i} \right),
    \label{eqn:rot-brownian}
\end{equation}
where the $\boldsymbol{\tau}$ quantities are torques and $\gamma_r$ is the rotational drag coefficient related to the corresponding rotational diffusion constant $D_r = \kT/\gamma_r$.
For our clathrin monomers, we use the values $D=13\,\unit{\micro\meter}^2/\mathrm{s}$ and $D_r = 0.03 \,\mathrm{rad}^2/\unit{\micro\second}$ as in previous studies\cite{guo2022large}.

\subsection{Over-Damped Helfrich Membrane Dynamics}

The most widely used meso-scale theoretical expression for the curvature-elastic free energy of a lipid bilayer is the Helfrich functional \cite{helfrich1973elastic}, in which the membrane is regarded as a two-dimensional manifold $\mathcal{M}$ with energy
\begin{equation}
    H = \int_\mathcal{M} \left( \frac{1}{2}\kappa_\mathrm{m} K^2 + \bar{\kappa}K_\mathrm{G} + \Sigma \right) \mathrm{d}A.
    \label{eqn:full-helfrich}
\end{equation}
Here $K$ is the sum of the local principal curvatures at a given point on the surface and $K_\mathrm{G}$ is their product (the Gaussian curvature).
The moduli $\kappa_\mathrm{m}$ and $\bar{\kappa}$ are the membrane bending rigidity and Gaussian rigidity, respectively, and $\Sigma$ is an isotropic surface tension.
In all that follows we will discard the energy arising from the Gaussian curvature term, as it is constant for topology-preserving deformations of membranes with no open edges \cite{deserno2015fluid,kreyszig1991differential}.

We will consider only membranes which exhibit small deviation away from flatness, so that we may employ the small-gradient approximation of the Helfrich energy functional. The membrane shape will be parametrized by its height $h(x,y)$ above some arbitrary base plane, in terms of which our small-gradient energy can be written as \cite{deserno2015fluid},
\begin{equation}
    E = \int_{\mathbb{R}^2} \left[ \frac{1}{2}\kappa_\mathrm{m} \left( \nabla^2 h \right)^2 + \frac{1}{2}\Sigma \left( \nabla h \right)^2 \right] \mathrm{d}x\mathrm{d}y.
    \label{eqn:small-grad-helfrich}
\end{equation}
We propagate the membrane in time using the Fourier-Space Brownian Dynamics (FSBD) method as elaborated by Lin and Brown \cite{lin2004brownian}, which has been used in numerous membrane simulation studies \cite{lepzelter2012integrin,badvaram2023physical,bihr2015multiscale,lin2005dynamic}, and which we briefly outline here.
The Monge-gauge equation of motion for a membrane whose energy is given by Eqn.~(\ref{eqn:small-grad-helfrich}) in the over-damped regime is \cite{seifert1997configurations,doi1988theory}
\begin{equation}
    \frac{\partial h(x,y)}{\partial t} = \int_{\mathbb{R}^2} \frac{1}{8\pi \eta |\mathbf{r}-\mathbf{r}'|}f(\mathbf{r}',t)\mathrm{d}x'\mathrm{d}y',
    \label{eqn:real-space-dynamics}
\end{equation}
where $f(\mathbf{r},t)$ is the local force per unit area in the $z$-direction, $\mathbf{r}=x\hat{\mathbf{x}} + y\hat{\mathbf{y}}$ is the position in the 2D plane, and $\eta$ is the viscosity of the embedding solvent.
The force $f$ includes both forces originating from the membrane energy $E$ as well as any external forces $f_\mathrm{ext}$ acting on the membrane,
\begin{equation}
    f(\mathbf{r},t) = -\frac{\delta E}{\delta h}(\mathbf{r},t) + f_{\mathrm{ext}}(\mathbf{r},t) + \xi(\mathbf{r},t).
\end{equation}
Here we have also included a term $\xi$ which accounts for a stochastic fluctuating force which thermalizes the membrane at a given temperature.
It is a zero-mean Gaussian which we will characterize further below.
The convolution over the entire membrane in Eqn.~(\ref{eqn:real-space-dynamics}) can be avoided by working in Fourier space.
We will consider a square membrane patch of dimension $L$ subject to periodic boundary conditions, for which our Fourier space representation is
\begin{equation}
    h(\mathbf{r},t) = \frac{1}{L} \sum_\mathbf{q} h_\mathbf{q}(t) e^{i \mathbf{q} \cdot \mathbf{r}},
\end{equation}
where $\mathbf{q}=\frac{2\pi}{L}(n\hat{\mathbf{x}} + m\hat{\mathbf{y}})$.
Plugging this into Eqn.~(\ref{eqn:real-space-dynamics}) results in a greatly simplified equation of motion for each Fourier mode:
\begin{equation}
    \frac{\partial h_{\mathbf{q}}}{\partial t} = \frac{f_{\mathbf{q}}}{4\eta q} = \frac{1}{4 \eta q} \Big[ -(\kappa_\mathrm{m} q^4 + \Sigma q^2)h_\mathbf{q} + f_{\mathrm{ext},\mathbf{q}} + \xi_\mathbf{q}  \Big],
    \label{eqn:fourier-eqn-motion}
\end{equation}
wherein we have replaced the membrane force $-\delta E / \delta h$ by its explicit Fourier space expression.
$f_{\mathrm{ext},\mathbf{q}}$ is the Fourier amplitude of the external force per unit area, which we acquire by Fast Fourier Transforming (FFT) all real-space forces arising from protein-membrane interactions at each timestep.
By requiring that $\xi_\mathbf{q}(t)$ satisfy the appropriate fluctuation-dissipation theorem, we can determine its power spectrum:
\begin{equation}
    \langle \xi_\mathbf{q}(t) \xi_\mathbf{q}^*(t') \rangle = 8 \eta q k_\mathrm{B}T \delta(t - t').
\end{equation}
We then integrate our equation of motion Eqn.~(\ref{eqn:fourier-eqn-motion}) using a simple Euler scheme with timestep $\Delta t$, wherein our random force at each time step is determined by integrating $\xi_\mathbf{q}$ over our finite timestep interval, yielding a Gaussian of variance $2\kT\Delta t/4\eta q$.

The coupling between the rigid-body proteins (whose propagation is handled by HOOMD-blue) and our Fourier-space membrane (propagated by our own code) is accomplished through the external force term in each corresponding equation.
For each spring connecting a protein to the membrane, the appropriate third-law force pairs are included in Eqns. (\ref{eqn:fourier-eqn-motion}), (\ref{eqn:com-brownian}), and (\ref{eqn:rot-brownian}).
In HOOMD-blue, this is accomplished by defining a custom force using \texttt{hoomd.md.force.Custom} to update the force arrays at each timestep.
Our continuum membrane is discretized in the numerical algorithm, where we represent the real-space membrane height with a square $N\times N$ grid of height values.
For each spring force $F$ on the membrane, we apply a force per unit area $f=F/(L/N)^2$ to the corresponding discrete patch of membrane over which the spring acts, as indicated in Fig.~\ref{fig:overview}a.
This discrete real-space representation of the force per unit area is Fourier transformed at each timestep using \texttt{SciPy}'s FFT library to acquire the amplitudes $f_{\mathrm{ext},\mathbf{q}}$ in Eqn.~(\ref{eqn:fourier-eqn-motion}).
Our implementation is available at \cite{code_repo}.

\subsection{Measuring Meso-Scale Rigidity via Buckling}

\begin{figure}
    \centering
    \includegraphics[width=\linewidth]{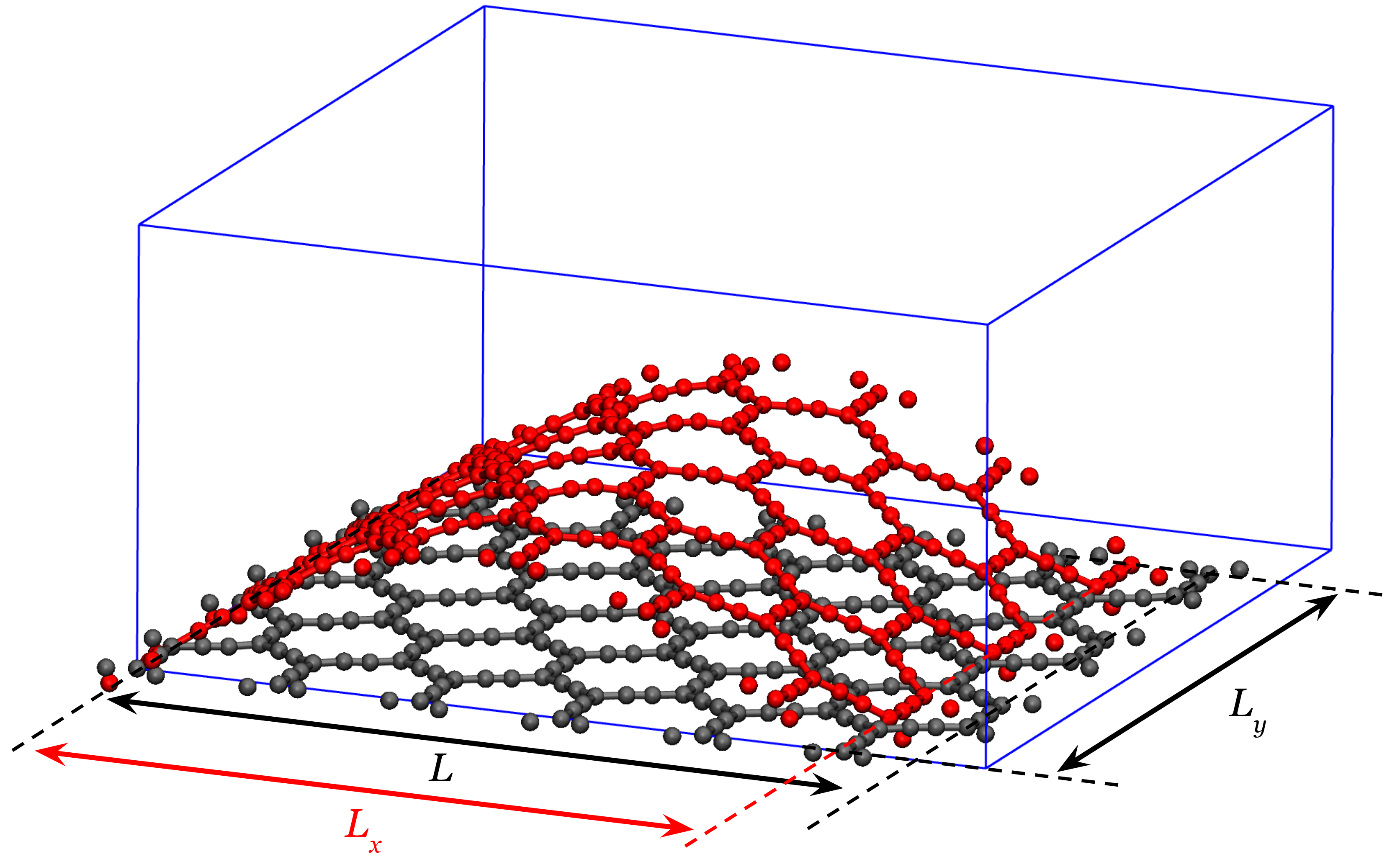}
    \caption{Example setup for the buckling protocol for measuring bending rigidity.
    Here the dimensionless buckling strain $\epsilon = (L-L_x)/L$ is 10 percent.}
    \label{fig:buckling}
\end{figure}

For sufficiently large clathrin coats deformed over sufficiently large length scales, one can model the deformation from the perspective of a thin two-dimensional sheet.
We thus model the elastic energy of the clathrin lattice when deformed away from its preferred curvature using a bending energy analogous to the Helfrich bending energy introduced for the membrane above \cite{tagiltsev2021nanodissected,walani2015endocytic,hassinger2017design}
\begin{equation}
    E_\text{clat} = \int \frac{1}{2}\kappa_{l} (K - K_0)^2 \mathrm{d}A,
    \label{eqn:clathrin-helfrich}
\end{equation}
where $\kappa_{l}$ is the protein lattice's bending modulus, $K$ is the sum of the coat's principal curvatures, and $K_0 = 2/R_0$ is its preferred curvature.
We will revisit the appropriateness of this choice of elastic energy in the results section.
The bending modulus $\kappa_{l}$ characterizes the meso-scale rigidity of the clathrin coat, and has been experimentally characterized to lie in the hundreds of $\kT$ range, though reported values vary \cite{tagiltsev2021nanodissected,nossal2001energetics,jin2006measuring}.
This will be our target meso-scale observable to which we will calibrate our CG model.
Since there are a range of values in the literature, here we will show how the bond parameters may be tuned to recapitulate any of the reported bending modulus values, rather than choosing to distinguish one experimental measurement above the others.
In order to compare our simulated coats to experiments, we must therefore measure the value of this elastic parameter in our CG models.

As a baseline calibration, we measured $\kappa_{l}$ using the buckling protocol \cite{hu2013determining,noguchi2011anisotropic}.
This method, previously used to characterize the bending modulus of lipid membranes in MD simulations \cite{hu2013determining,diggins2015curvature}, determines the bending modulus of a thin sheet by measuring its stress-strain relation when buckled under a lateral load.
The rigidity for this mode of bending is typically called the ``flexural rigidity'' \cite{landau2012theory}, and as such we will denote its value by $\kappa_{l,\mathrm{flex}}$ to distinguish it from rigidities measured via our second protocol which we introduce later.
We prepare our flexible clathrin coat (with pucker angle $\alpha=90^\circ$, thus $K_0=0$) in a flat rectangular configuration with length $L$ and then simulate it buckled (forced into a projected length $L_x < L$) at successively higher strains (smaller $L_x$), measuring the equilibrium force required to maintain it at the given strain (see Fig.~\ref{fig:buckling}).
The bending modulus is determined by fitting this force-vs.-strain data to a theoretical curve derived from the energy in Eqn.~(\ref{eqn:clathrin-helfrich}), expressed as a power series in the dimensionless buckling strain $\epsilon = (L - L_x)/L$,
\begin{align}
    F_x(\epsilon) &= \kappa_{l,\mathrm{flex}} L_y \left( \frac{\pi}{L} \right)^2 \sum_{n=0}^\infty b_n \epsilon^n
    \label{eqn:buckling_force}
    \\
    &= \kappa_{l,\mathrm{flex}} L_y \left( \frac{\pi}{L} \right)^2 \left[ 1 + \frac{1}{2} \epsilon + \frac{9}{32} \epsilon^2 + \frac{21}{128} \epsilon^3 + \cdots \right]
    \nonumber
\end{align}
where $F_x$ here is the force in the buckling direction and the coefficients $b_n$ are given up to 10\textsuperscript{th} order in ref \cite{hu2013determining}. This allows for a simple one-parameter fit to determine $\kappa_{l,\mathrm{flex}}$ (the numerical prefactor in Eqn.~(\ref{eqn:buckling_force}) differs slightly from the one presented in ref. \cite{hu2013determining} due to altered boundary conditions; see derivation in SI Sec. 1).

For the results presented in the main text, all buckling strains were imposed along the $\hat{x}$-direction as shown in the figure.
Buckling along the $\hat{y}$-direction is technically distinct from this, as the hexagonal lattice is not rotationally symmetric.
A small series of calibration simulations showed a modest deviation between $\kappa_{l,\mathrm{flex}}$ values measured from buckling along $\hat{y}$ and $\hat{x}$ (roughly 10\% over the regime tested, see Fig. S2 in the SI), but with the same scaling behavior.

\subsection{Effective Lattice Rigidity Measured from Membrane Deformation Induced by a Spherical Clathrin Coat}

\begin{figure}
    \centering
    \begin{overpic}[width=\linewidth]{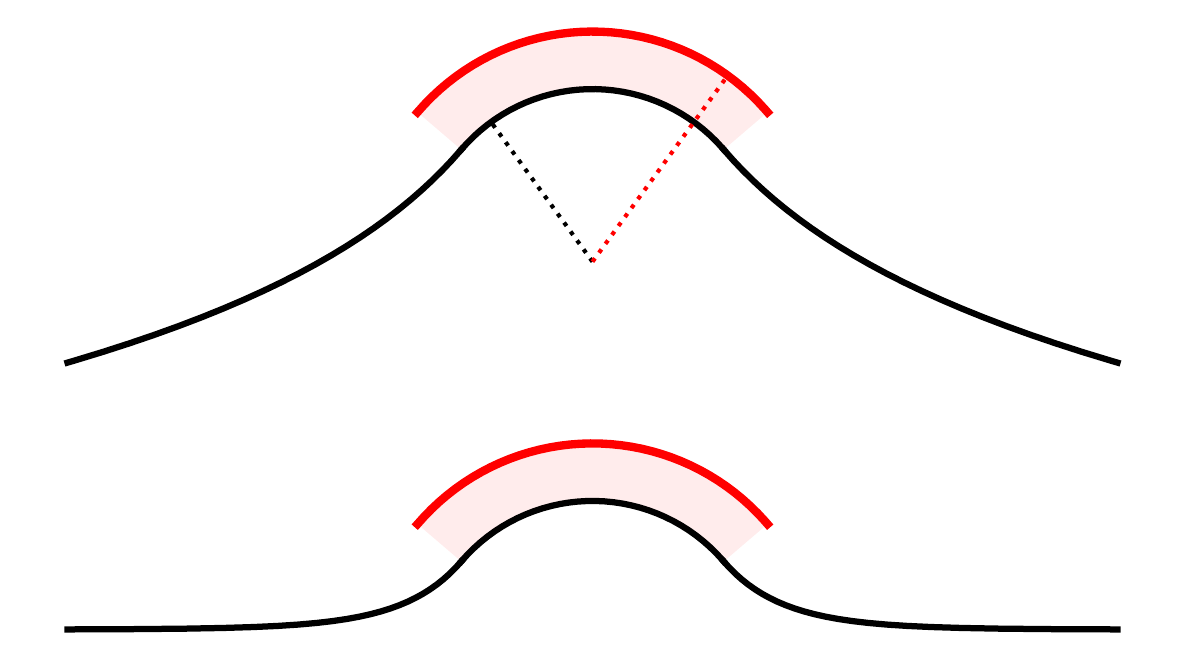}
        \put(36,35){\small $R-\ell$}
        \put(55,37){\small \textcolor{red}{$R$}}
        \put(60,52){\small \textcolor{red}{Area $a$}}
        \put(75,33.5){\small $\Sigma = 0$}
        \put(75,4){\small $\Sigma > 0$}
    \end{overpic}
    \caption{Top: Cross-section of a partial SCC (red) deforming a tensionless membrane (black).
    The curvature radii for the coat and adhered portion of the membrane are indicated by the dotted lines of the corresponding color.
    The region of adhesion is shaded between the SCC and the membrane.
    Bottom: The same as above but for a non-zero membrane tension, resulting in an asymptotically flat height profile.}
    \label{fig:theory-diagram}
\end{figure}

Since clathrin assembles into approximately spherical basket structures to perform its endocytic function, here we give an approximate treatment of curvature induction by a partial spherical clathrin coat (SCC). This also provides an alternate way to infer a continuum rigidity $\kappa_{l}$ for our clathrin lattices, by quantitatively measuring the extent to which they are able to deform the (continuum-model) membrane in simulation.
Fig.~\ref{fig:theory-diagram} shows a simplified diagram of a protein coat of area $a$ which has deformed a tensionless membrane.
For our derivation we assume that the membrane is perfectly offset from its adherent protein coat by a distance $\ell$ (accounting for the presence of clathrin adaptors \emph{in vivo}), such that the membrane in this region has radius of curvature $R-\ell$ over a smaller area $a_\mathrm{m} = a (1 - 2\ell/R + \ell^2/R^2)$. Since $(R-\ell)^2$ and $a_\mathrm{m}$ have the exact same $\ell$-scaling, the membrane bending energy in this region is analytically identical to the case where $\ell$ is neglected. Lastly, in the absence of tension, the non-adhered portion of the membrane assumes the form of a minimal surface for which $K=0$, and thus accumulates no energy. This leaves us with
\begin{equation}
    E = \left[\frac{1}{2}\kappa_\mathrm{m} K^2 + \frac{1}{2}\kappa_{l}\left( K - K_0 \right)^2 \right] a,
\end{equation}
whose minimization yields the equilibrium SCC curvature
\begin{equation}
    K_\mathrm{eq} = \frac{\kappa_{l}}{\kappa_\mathrm{m} + \kappa_{l}} K_0.
    \label{eqn:eq-K}
\end{equation}
Since we control $\kappa_\mathrm{m}$ in simulation and can readily measure $K_0$ from simulations of non-adhered free lattices, this zero-tension case gives us a simple one-parameter fit for the effective clathrin rigidity $\kappa_{l}$ by simulating partially-assembled spherical clathrin lattices adhered to membranes of varying rigidity and measuring their resulting curvature $K_\mathrm{eq}$.
We will refer to the rigidity inferred this way as the ``effective bending rigidity'', $\kappa_{l,\mathrm{eff}}$.

Membrane tension is another parameter which will influence the curvature state of both the membrane and adhered protein coat.
The presence of a constant membrane tension $\Sigma$ introduces a term $\Sigma \Delta A$ to the free energy, where $\Delta A$ is the difference between the membrane area when adhered and the corresponding flat area \cite{deserno2004elastic}.
A simple geometric calculation in which it is assumed that the adhered membrane and coat have the same radius (neglecting $\ell$) yields $\Delta A = a^2/4 \pi R^2$ for the portion of the membrane that is adhered.
With sufficient tension, the rest of the membrane will be approximately flat (Fig.~\ref{fig:theory-diagram}), and so we will simply take
\begin{equation}
    E \approx \left[\frac{1}{2}\kappa_\mathrm{m} K^2 + \frac{1}{2}\kappa_{l}\left( K - K_0 \right)^2 + \frac{\Sigma a}{4\pi R^2} \right] a
\end{equation}
as our new energy.
Minimizing with respect to $K=2/R$ then yields
\begin{equation}
    K_\mathrm{eq}(\Sigma) \approx K_0 \left[ 1 + \frac{1}{\kappa_{l}}\left( \kappa_\mathrm{m} + \frac{\Sigma a}{8 \pi} \right) \right]^{-1}.
    \label{eqn:Keq-simple}
\end{equation}
We thus see that in this limit, increasing the membrane tension $\Sigma$ has the result of effectively stiffening the membrane, as its contribution is added directly to $\kappa_\mathrm{m}$.
We must note that the result above is only reasonable in the limits of $\ell/R \ll 1$ and $a/2\pi R^2 \ll 1$, and a more thorough treatment of tension must explicitly account for changes in membrane shape in the non-adhered region.
The rigorous result still gives an energy contribution that scales approximately as the product $\Sigma a^2$, allowing us to take the above as a reasonable simplified result; see  SI Sec. 6 for details.

\subsection{Setup and Parameters}

The timestep for all membrane-protein interaction simulations was set to $\Delta t = 0.005 \mathrm{ns}$.
For the membrane dynamics we use a solvent viscosity $\eta = 24 \, \kT \, \mathrm{ns}\, \mathrm{nm}^{-3}$, approximating the cytosolic viscosity used in Ref.~\cite{lin2004brownian} at $T=310$K.
Unless noted otherwise, all membrane simulations were carried out over a square base plane of dimension $L=200\mathrm{nm}$ with $N=55$ subdivisions in each direction for a grid size of approximately $3.64\mathrm{nm}$ subject to periodic boundary conditions.
This $N$ is chosen such that the grid length scale is comparable to typical membrane thickness values.
For an over-damped, tensionless Helfrich membrane, the relaxation timescale for an undulation with wave-vector $\mathbf{q}$ is $\tau_\mathbf{q} = 4\eta / \kappa q^3$, as can be seen from Eqn.~(\ref{eqn:fourier-eqn-motion}).
The slowest mode is thus the one with the smallest wave-vector (largest wavelength), so we calculate for $q_\mathrm{min}=2\pi/L$ that we have $\tau_{\mathbf{q}} \approx 310 \unit{\micro\second}$.
With this in mind, 
the total simulation time was 1 ms, of which the first 0.5 ms were discarded for equilibration.
If not stated, the membrane bending modulus is $\kappa_\mathrm{m}=20\,\kT$, and we assume a membrane with zero spontaneous curvature throughout.
For the clathrin lattice simulations, all protein-membrane bonds have an equilibrium length of $\Delta z = 15\mathrm{nm}$ and bond stiffness $k_\mathrm{mem} = 1000 \kT/\mathrm{nm}^2$ unless specified otherwise.
All spherical clathrin coat simulations presented use a pucker angle $\alpha=107.5^\circ$.
Buckling simulations were carried out using the \texttt{ESPResSoMD} package version 4.2.1 \cite{weik2019espresso}. All other simulations were carried out using HOOMD-blue version 5.3.1\cite{anderson2020hoomd}.
Code for carrying out all types of simulations presented here are publicly available at \cite{code_repo}. We include a representative simulation trajectory in Movie S1. 

\section{Results}

\subsection{$k_\theta$ Controls Flexural Rigidity of Clathrin Coat}

\begin{figure}
    \centering
    \begin{overpic}[width=\linewidth]{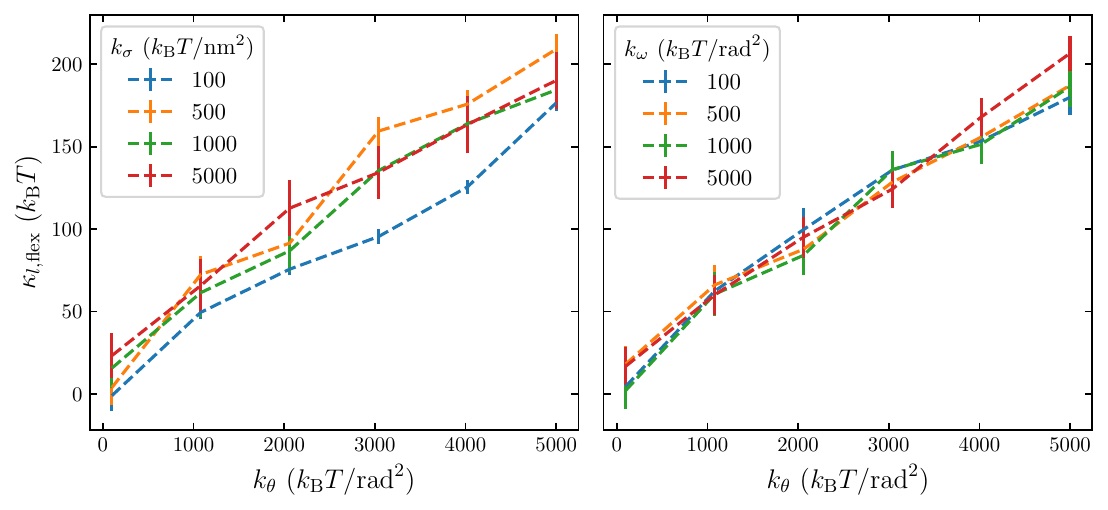}
        \put(46,10){\textbf{a} } 
        \put(92,10){\textbf{b} }
    \end{overpic}
    \caption{Flexural bending rigidity $\kappa_{l,\mathrm{flex}}$ measured from buckling simulations (see Fig.~\ref{fig:buckling}) where the most critical bond rigidity $k_\theta$ is varied along the $x$-axis. The other two bond rigidities are varied from low (blue) to high (red) rigidities of a) $k_\sigma$ and b) $k_\omega$. In each plot, the parameter which does not appear in the legend has been averaged over using a simple unweighted arithmetic average. The un-averaged data are presented in Fig. S3 of the SI. Error bars give the error of the mean determined by standard error propagation of the individual simulation errors, which were determined by blocking analysis\cite{flyvbjerg1989error,foley2024quantifying} (See SI Sec. 4 for details of the blocking procedure).}
    \label{fig:buckleresults}
\end{figure}

Measurements of the bending rigidity $\kappa_{l}$ from buckling simulations of flat ($\alpha=90^\circ$) clathrin lattices are summarized in Fig.~\ref{fig:buckleresults}. The lattices were assembled from 72 individual clathrin triskelia as depicted by the grey initial clathrin sheet in Fig.~\ref{fig:buckling}.
To systematically assess the role of the separate microscopic bond parameters $k_\sigma$, $k_\theta$, and $k_\omega$, we performed a parameter sweep, running a total of 96 buckling simulations for all combinations of parameters where the individual constants take on the following values; $k_\sigma$: 100, 500, 1000, and 5000 $\kT/\mathrm{nm}^2$, $k_\omega$: 100, 500, 1000, and 5000 $\kT/\mathrm{rad}^2$, $k_\theta$: 100, 1080, 2060, 3040, 4020, and 5000 $\kT/\mathrm{rad}^2$.

The clear trend in these buckling simulations is that $k_\theta$ controls $\kappa_{l,\mathrm{flex}}$, with approximately linear proportionality.
This makes intuitive sense, as the bond angles $\theta_1$ and $\theta_2$ are the geometric parameters principally responsible for bending the overall structure.
The value of $k_\sigma$ plays a small role in this rigidity in that for sufficiently small values, the clathrin sheet may relieve buckling strain by compressing bond lengths rather than through bending.
Above some small value of $k_\sigma$, however, no further effect is observed within statistical error (Fig.~\ref{fig:buckleresults}a).
The torsional parameter $k_\omega$ has no discernible effect on the measured bending rigidity (Fig.~\ref{fig:buckleresults}b).

\subsection{Spherical Clathrin Coats Have Much Larger Effective Rigidity}

\begin{figure*}
    \centering
    \begin{minipage}{0.55\linewidth}
        \begin{overpic}[width=\linewidth]{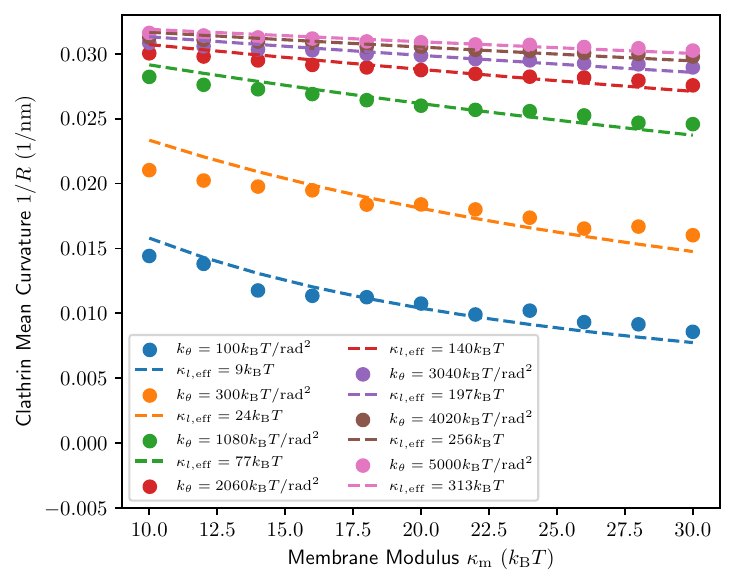}
            \put(75,13){\includegraphics[width=0.2\linewidth]{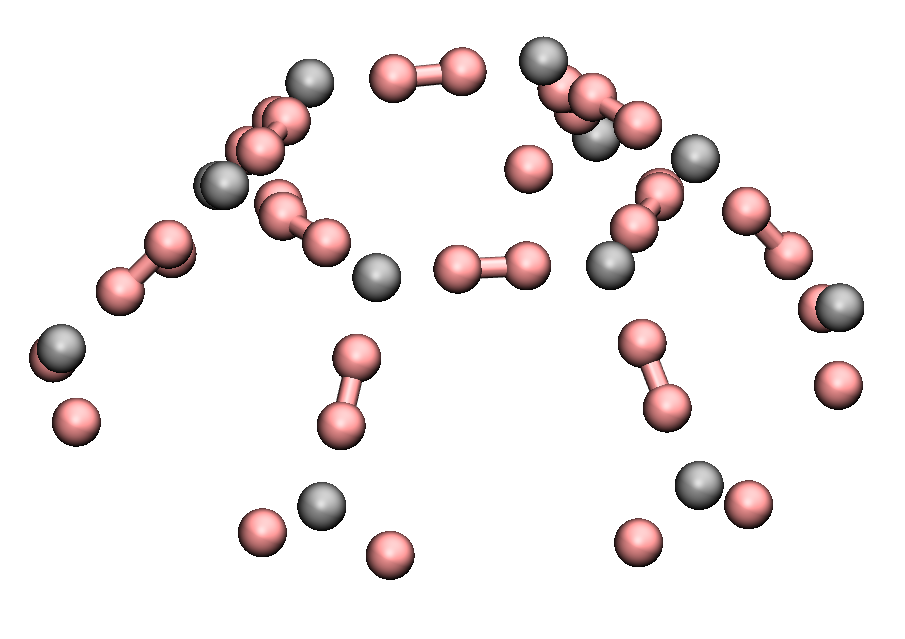} }
            \put(2,74){\Large \textbf{a}}
            \put(104,72){\Large \textbf{b}}
        \end{overpic}
    \end{minipage}
    \begin{minipage}{0.36\linewidth}
            \hfill
            \includegraphics[width=0.42\linewidth]{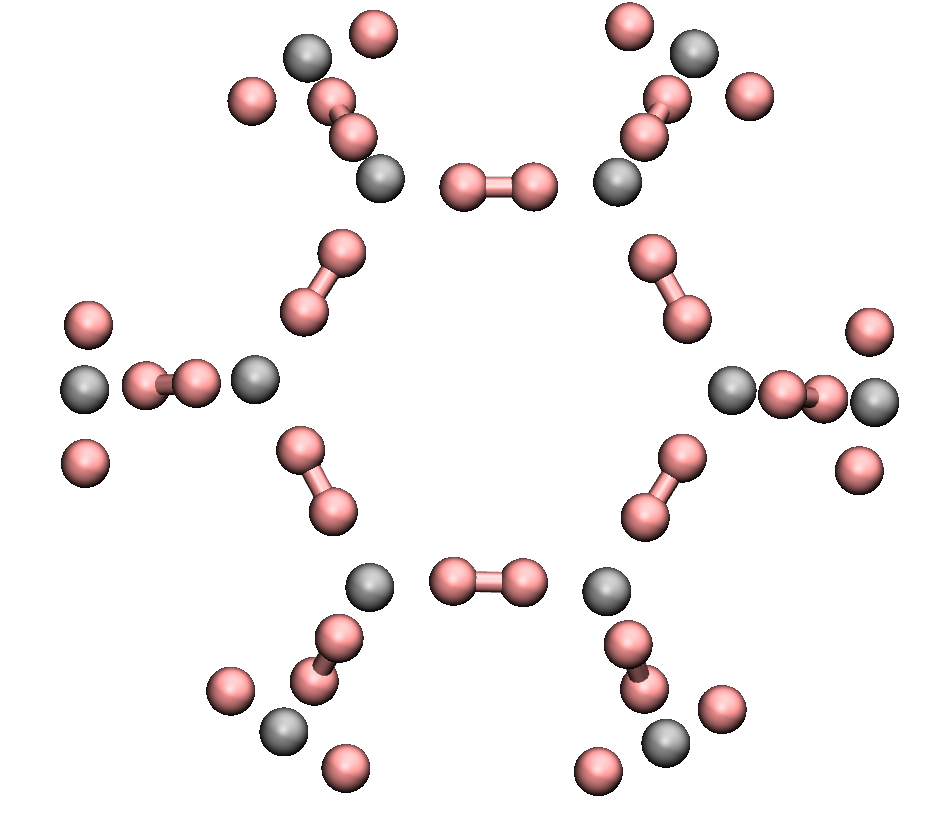}
            \includegraphics[width=0.42\linewidth]{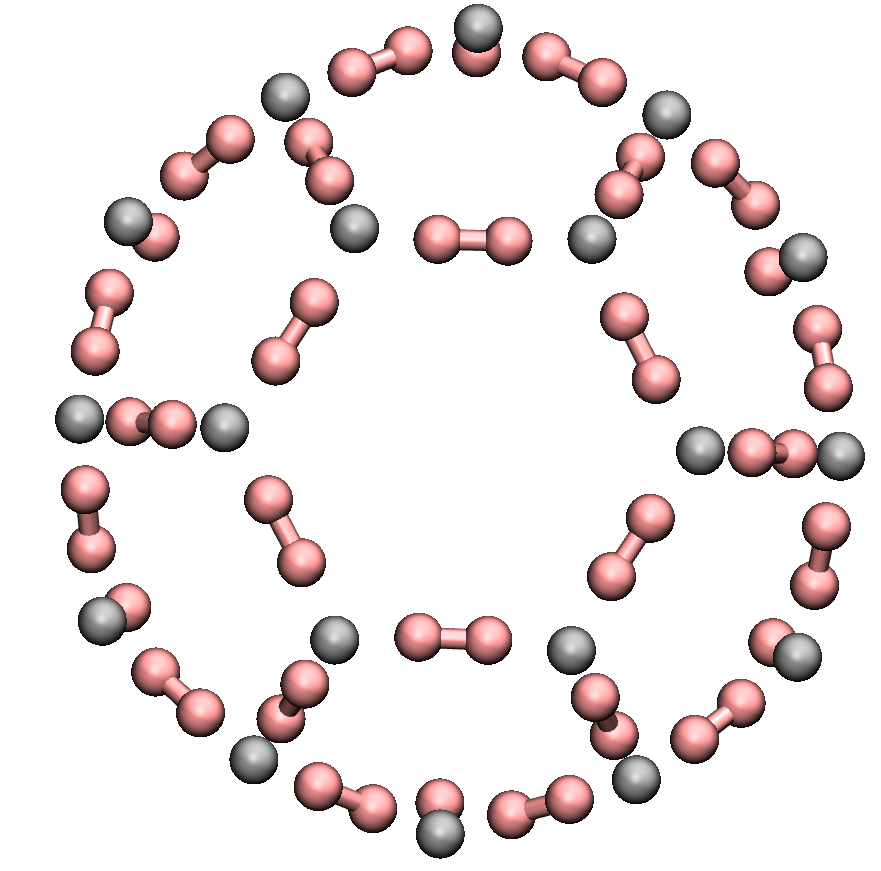}
            \\
            \begin{overpic}[width=\linewidth]{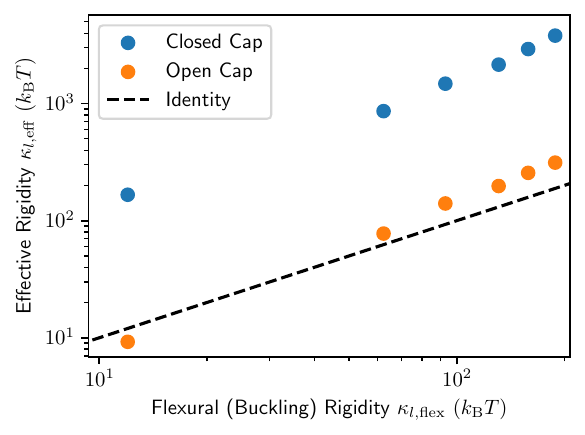}
                \put(4,67){\Large \textbf{c}}
            \end{overpic}
        \end{minipage}
    \caption{
        (\textbf{a})
        Plots of measured mean curvature $H=1/R$ vs. membrane bending modulus $\kappa_\mathrm{m}$ for varying values of bond parameter $k_\theta$ for $\alpha=107.5^\circ$ open cap SCCs. Dashed curves are fits to Eqn.~(\ref{eqn:eq-K}) with corresponding best fit value of $\kappa_{l}$ given in the legend.
        (\textbf{b}) Top-down view of the two CG SCC structures simulated. Left: 12-mer open cap. Right: 18-mer closed cap. 
        (\textbf{c}) Plot of inferred effective spherical rigidity vs. flexural rigidity measured via buckling for the same values of microscopic bond parameters $k_x$. The membrane tension $\Sigma$ was set to zero for all spherical cap simulations which generated these data. The identity line shows where points would fall if both methods yielded identical rigidity values.
    }
    \label{fig:sphere-vs-buckle}
\end{figure*}

We carried out a suite of simulations for two distinct semi-spherical clathrin structures (shown in Fig.~\ref{fig:sphere-vs-buckle}b).
Both are sub-structures of the ``D6 barrel'' 36-mer clathrin basket observed in natively assembled endocytic vesicles \cite{morris2019cryo,paraan2020structures}.
The first is a complete hexameric ring of clathrin monomers with an additional clathrin monomer bonded to each of the central six, creating a 12-mer which we refer to as the ``open cap'' structure (Fig.~\ref{fig:sphere-vs-buckle}b, left).
Our second structure, which we call the ``closed cap'', is an 18-mer formed by adding 6 more clathrin monomers to the open cap, which completes the next geometric ring around the central hexamer resulting in a hexagon surrounded by six pentagonal faces (Fig.~\ref{fig:sphere-vs-buckle}b, right).
These can be seen as a possible early assembly intermediate along the way to the complete D6 barrel structure.
In order to generate a CG model more representative of one which would natively assemble into this shape, we simulated the 18-mer structure on its own over a range of pucker angle values $\alpha$ and chose the one which yielded the smallest potential energy, $\alpha=107.5^\circ$ (see Fig. S6 in the SI).
The measured equilibrium cage radius in the free simulation was used to determine the spontaneous radius of curvature $R_0 = 2/K_0$ for the continuum clathrin energy expression Eqn.~(\ref{eqn:clathrin-helfrich}), and its value varied less than $0.1 \mathrm{nm}$ across our range of $k_\theta$ values.

For both structures, we carried out simulations in which they are tethered to our FSBD membrane model at zero tension and varied $k_\theta$ over the same set of values that were used for the buckling simulations (see III.A and Fig.~\ref{fig:buckleresults} above).
$k_\sigma$ and $k_\omega$ were fixed at $500 \kT/\mathrm{nm}^2$ and $500 \kT/\mathrm{rad}^2$, respectively.
We carried out simulations at 11 different membrane bending modulus $\kappa_\mathrm{m}$ values evenly spaced from $10\kT$ to $30\kT$ for both structures.
From these simulations we extracted the average curvature radius of the coat as determined by a best-fit sphere.
From these measured curvatures, the best-fit value of $\kappa_{l,\mathrm{eff}}$ was determined via fits to Eqn.~(\ref{eqn:eq-K}).
The results, which are strikingly different for the two geometries, are shown in Fig.~\ref{fig:sphere-vs-buckle}.
When the coat is less rigidly constrained into the spherical shape, as in the open cap configuration, the measured $\kappa_{l,\mathrm{eff}}$ is very nearly in agreement with the flexural rigidity $\kappa_{l,\mathrm{flex}}$ measured via buckling.
However, the inferred value of $\kappa_{l,\mathrm{eff}}$ found from the closed cap simulations is more than an order of magnitude larger in all cases.

To understand this discrepancy, we must revisit the assumptions that underlie the continuum energy Eqn.~(\ref{eqn:clathrin-helfrich}) used for the clathrin lattice. We have modeled the clathrin coat using only a Helfrich-like bending energy which quadratically penalizes mean curvature deviation from the coat's preferred curvature, a modeling strategy has been used in several studies \cite{frey2024coat,tagiltsev2021nanodissected,walani2015endocytic,hassinger2017design}.
However, this approximation leaves out other elastic energy contributions, altering the interpretation of the elastic parameters in the meso-scale continuum model.
The energy functional given in Eqn.~(\ref{eqn:clathrin-helfrich}) is exact (up to quadratic order) for \emph{fluid membranes}, for which the in-plane shear modulus is zero and constituent molecules (lipids) are free to rearrange themselves.
Clathrin lattices, and other solid-like thin-sheet assemblies, are fundamentally different primarily in that their bond network provides shear rigidity, which dramatically changes the bending energetics for deformations beyond pure bending \cite{roos2010physical}.
For pure bending, the elastic energy does in fact have a form that exactly matches Eqn.~(\ref{eqn:clathrin-helfrich}), and the ``bending modulus'' $\kappa_{l}$ that appears is properly referred to as the \emph{flexural} or \emph{cylindrical} rigidity \cite{landau2012theory}, as introduced earlier.

However, deformations other than pure cylindrical bending, as in the case of spherical shell deformations, do not generally have a Helfrich-like energy \cite{roos2010physical}.
When deforming a spherical shell, stretching enters at lowest order along with bending, unlike in the case of pure (cylindrical) bending \cite{landau2012theory,roos2010physical}.
Since our lattice shape remains nearly spherical for our simulated SCCs, we can of course still write down a local energy density as a quadratic-level expansion with respect to our main shape parameter (curvature, or equivalently the radius of the cap), and which therefore takes a form that matches Eqn.~(\ref{eqn:clathrin-helfrich}).
However, due to the fundamentally different nature of these deformations, the effective bending modulus $\kappa_{l,\mathrm{eff}}$ inferred from this energy will generally be radically different from that measured for pure bending, as in our buckling measurements above.
As such, the Helfrich-like energy function is still useful for characterizing the effective rigidity of a spherical clathrin lattice.
However, care must be taken in the parameter interpretation, because the modulus associated with this energy is \emph{geometry-dependent}, as seen in Fig.~\ref{fig:sphere-vs-buckle}c, and not generally applicable to other modes of deformation or different structures with the same underlying microscopic parameters.

We also note here that there appears to be a small systematic deviation between the observed SCC curvature and the theoretical model visible in Fig.~\ref{fig:sphere-vs-buckle}a.
The theory curve is consistently slightly above the measurements at low $\kappa_\mathrm{m}$ and slightly below for high $\kappa_\mathrm{m}$.
We believe this to be a compounded result of multiple approximations made in the derivation of Eqn.~(\ref{eqn:eq-K}): (\textit{i}) The prediction was derived from the full non-linear Helfrich model, whereas the simulated membrane uses the small-gradient approximation, (\textit{ii}) the derivation assumes the protein coat has \textit{constant} area $a$, which is not true in simulation, and (\textit{iii}) the theory assumes uniform adhesion in a perfectly spherical shape, neither of which is true in the simulations.
While addressing all of these details would likely give better agreement between theory and simulation, we doubt that the significantly increased analytical complexity would yield much difference in the fitted value of $\kappa_{l,\mathrm{eff}}$.

\subsection{SCC Deformation under Membrane Tension is Consistent with Theory}

\begin{figure*}
    \centering
    \begin{minipage}{0.53\linewidth}
        \begin{overpic}[width=\linewidth]{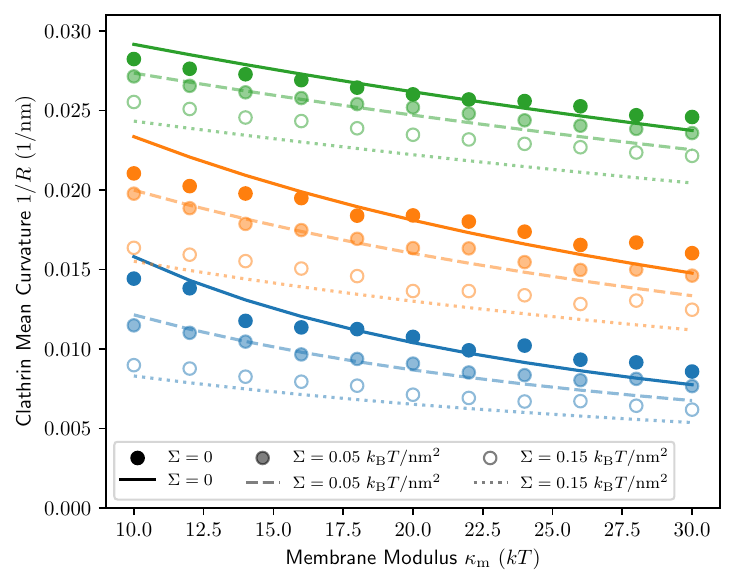}
            \put(2,74){\large \textbf{a}}
        \end{overpic}
    \end{minipage}
    \begin{minipage}{0.42\linewidth}
        \hfill
        \begin{overpic}[width=0.93\textwidth]{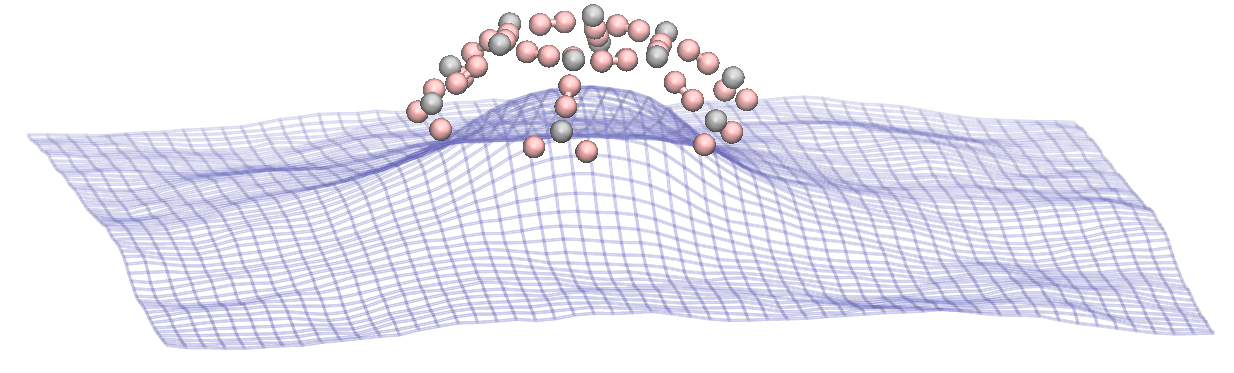}
            \put(3,28){\large \textbf{b}}
        \end{overpic}
        \\
        \begin{overpic}[width=\textwidth]{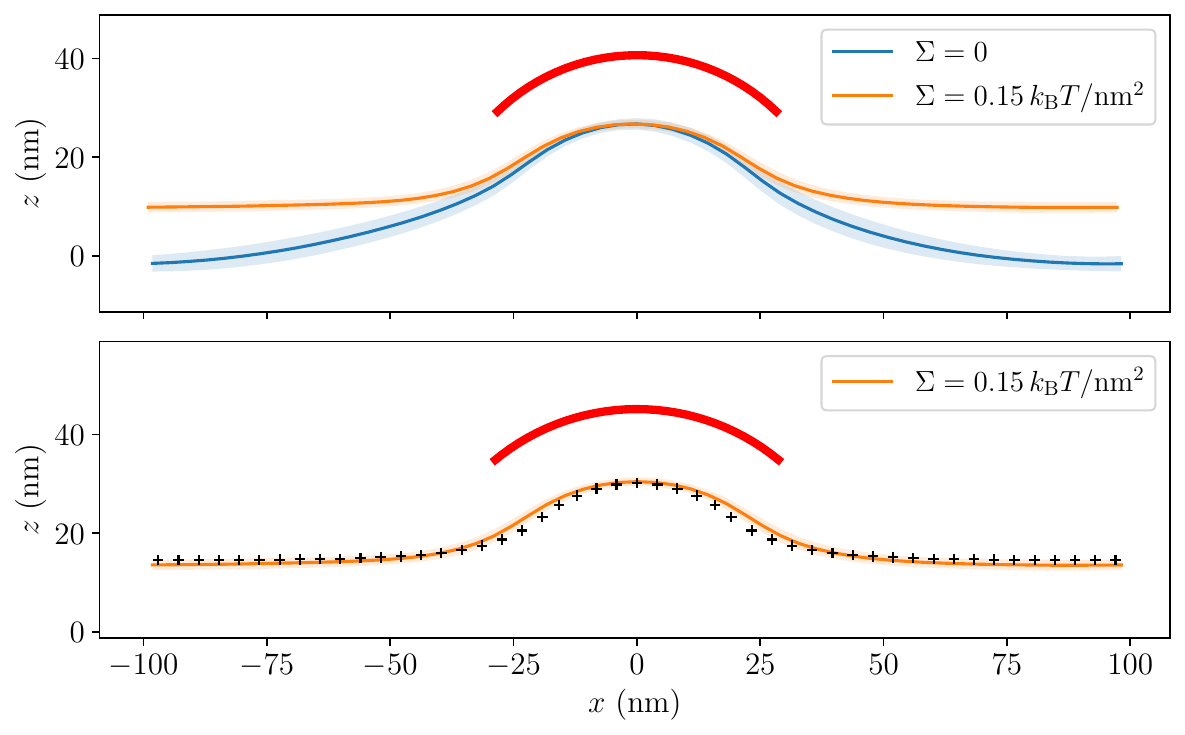}
            \put(10,55){\large \textbf{c}}
            \put(10,27){\large \textbf{d}}
        \end{overpic}
    \end{minipage}
    \caption{
        (\textbf{a}) Plots of mean curvature of SCC measured from $\alpha=107.5^\circ$ open cap simulations with varying membrane tension $\Sigma$ for coat rigidities $\kappa_{l,\mathrm{eff}}=9\,\kT$   (blue),$\, 24\,\kT$ (orange), and $77\,\kT$ (green). Points are simulation data, solid curves are the same theory fits from Fig.~\ref{fig:sphere-vs-buckle}a. The dashed and dotted curves are predictions of the SCC curvature using Eqn.~(\ref{eqn:Keq-simple}) based on the corresponding zero-tension fit. 
        (\textbf{b}) Snapshot of a $\Sigma=0$ open cap simulation with $k_\theta=1080 \,\kT/\mathrm{rad}^2$.
        (\textbf{c}) Average membrane height profile (blue and orange) along a slice through the membrane centered underneath the SCC for two simulations: blue, the same simulation as in \textbf{b}; orange, the same system but with membrane tension $\Sigma=0.15\,\kT/\mathrm{nm}^2$. The red arc corresponds to the average measured location and curvature of the SCC in the zero-tension simulation. Shading indicates the standard deviation of observed height.
        (\textbf{d}) The same $\Sigma=0.15\,\kT/\mathrm{nm}^2$ system as \textbf{c}, with a comparison to the shape profile predicted by Eqn.~(\ref{eqn:exact-hr}), shown as the + symbols.
    }
    \label{fig:tension-R}
\end{figure*}

To assess the degree to which our clathrin-membrane simulations agree with meso-scale theoretical results, we also simulated the 12-triskelion open cap adhered to membranes (see Fig.~\ref{fig:tension-R}b) with tension values up to $\Sigma=0.15\,\kT/ \mathrm{nm}^2$, which range from typical to large values of cellular membrane tensions \cite{morris2001cell,kozlov2015membrane}.
We once again measured the average radius of curvature via best-fit sphere, and the results (for $\Sigma=0,$ $0.05$, and $0.15\,\kT/\mathrm{nm}^2$) are plotted in Fig.~\ref{fig:tension-R}a.
Also plotted in this figure are the theoretical predictions for the equilibrium curvature radius as calculated from Eqn.~(\ref{eqn:Keq-simple}).
For the evaluation of Eqn.~(\ref{eqn:Keq-simple}), we used the value of $\kappa_{l,\mathrm{eff}}$ as determined from the zero-tension fits in Fig.~\ref{fig:sphere-vs-buckle}a, and estimated the value of the coat area $a$ from the equilibrium average of the simple polygon area of the CG structure in simulation.
The agreement between the predicted curvature deviation under tension and the simulation measurements is good, although the high tension systems seem to have slightly higher curvature than predicted by the simplified theory.
The cause of this discrepancy is likely related to points \textit{ii} and \textit{iii} discussed in the final paragraph of Sec. III.B, as well as the limited validity of the approximation made in the derivation of Eqn.~(\ref{eqn:Keq-simple}), as discussed in the final sentences of Sec. II.D.

Figures \ref{fig:tension-R}c and \ref{fig:tension-R}d plot the observed average membrane profile along a slice that passes directly under the center of the open cap in simulation.
In Fig. \ref{fig:tension-R}d we additionally plot the exact analytical solution of the small-gradient Helfrich shape equation for a membrane adhered to a spherical shape of radius $R_\mathrm{m}=R-\ell$, where the free (non-adhered) portion of the membrane has its height $h(r)$ as a function of radial distance $r$  given by \cite{deserno2004elastic}
\begin{equation}
    \frac{h(r)}{R_\mathrm{m}} =  1 - z - \frac{\lambda}{R_\mathrm{m}} \frac{k}{1 - z} \frac{K_0(k R_\mathrm{m} / \lambda) - K_0(r / \lambda)}{K_1(k R_\mathrm{m} / \lambda)},
    \label{eqn:exact-hr}
\end{equation}
where $K_n(x)$ are modified Bessel functions of the second kind, $z = a/2 \pi R^2$, $k=\sqrt{z(2-z)}$, and $\lambda=\sqrt{\kappa_m/\Sigma}$ is the membrane relaxation length scale.
Despite our membrane not being rigidly adhered to a spherical shape (see Fig. S7 in the SI), but rather harmonically bonded at discrete points to a flexible lattice, we find very good agreement between the analytical and simulated shape profiles for $\Sigma=0.15\,\kT/\mathrm{nm}^2$.
In the limit $\Sigma\rightarrow 0$ ($\lambda \rightarrow \infty$), the membrane no longer becomes asymptotically flat, and $h(r)$ for a free membrane becomes logarithmic (in the small-gradient Helfrich approximation).
As such, we are unable to compare our $\Sigma=0$ profile to this theoretical expression due to the influence of the periodic boundary conditions, which require the membrane to reach $h'(r)=0$ within the finite range of our periodic simulation box.

\subsection{Lattice Geometry and Membrane Tension Affect Lattice Relaxation Timescale}

\begin{figure}
    \centering
    \begin{overpic}[width=0.48\linewidth,trim=0.5em 0 0.5em 0]{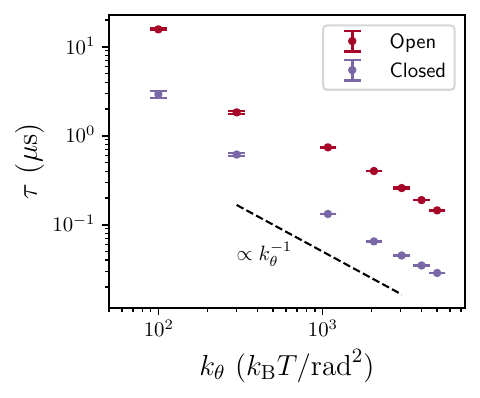}
        \put(25,23){\textbf{a}}
    \end{overpic}
    \begin{overpic}[width=0.48\linewidth,trim=0.5em 0 0.5em 0]{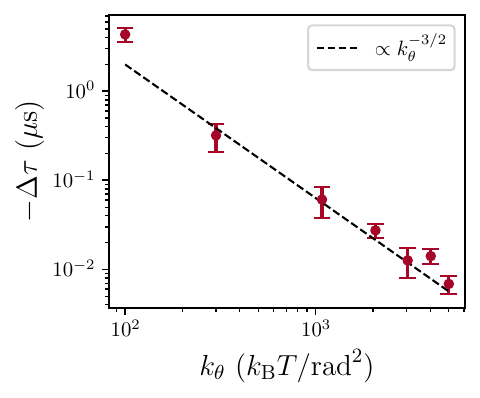}
        \put(25,23){\textbf{b}}
    \end{overpic}
    \caption{
        (\textbf{a}) Curvature relaxation time $\tau$ \textit{vs.} $k_\theta$ for open and closed cap lattices at zero tension.
        Dashed line shows $\propto k_\theta^{-1}$ power law scaling.
        (\textbf{b}) $-\Delta \tau = \tau(0) - \tau(\Sigma)$ \textit{vs.} $k_\theta$ for $\Sigma=0.15\,\kT/\mathrm{nm}^2$, showing that the relaxation timescale is decreased when membrane tension is applied, with the effect being larger for weaker lattices. The dashed line again shows a power law scaling, this time with exponent $-3/2$.
        See SI Sec. 5 for details on the error bars in these plots.
    }
    \label{fig:relaxation-time}
\end{figure}

To assess the impact of lattice rigidity and membrane tension on the dynamics of the system, we extracted the curvature relaxation timescale from each SCC simulation trajectory.
In all cases, the membrane is initially flat. The protein subunits are individually puckered (Fig 1), but the lattice is placed such that all subunit COM are within a flat plane at the equilibrium separation from the membrane for the open cap. For the closed cap, the COM of the central hexagon is shifted up by 5 nm initially to prevent inversion. This lattice still transitions initially to a flat state due to the slower hydrodynamic mode of the membrane (Movie S2).
For each trajectory, we fit the instantaneous measured SCC curvature $K(t)$ to an exponential decay to the equilibrium value,
\begin{equation}
    K(t) = K_{\mathrm{eq}}\left( 1 - e^{-t/\tau} \right),
    \label{eqn:relaxation-kinetics}
\end{equation}
where $\tau$ is the fitted relaxation timescale.
Fig. S4 of the SI shows an example fit.
The results are summarized in Fig.~\ref{fig:relaxation-time}.
In Fig.~\ref{fig:relaxation-time}a we compare the curvature relaxation timescale of the open and closed cap lattices at zero tension, wherein we see that the timescales are substantially lower for the closed cap, meaning faster relaxation dynamics.
The relaxation timescale appears to have approximate $\propto k_\theta^{-1}$ power-law scaling for both the open and closed cap lattices, as might be expected for over-damped dynamics, producing faster, sub-$\mu$
s relaxation with stiffer lattices. 
In Fig.~\ref{fig:relaxation-time}b, we examine how the relaxation timescale changes for the open cap when a membrane tension of $\Sigma=0.15\,\kT/\mathrm{nm}^2$ is applied.
In all cases, the curvature relaxes faster ($\Delta \tau < 0$, hence plotting $-\Delta \tau$ in log-log).
Interestingly, the timescale difference $\Delta \tau$ also approximately follows a power law $\propto k_\theta^{-3/2}$, though it is hard to say whether this is a true fractional power law or a crossover between two integer powers.

\subsection{Transfer of Arbitrary Structures from NERDSS}

\begin{figure*}
    \centering
    \begin{overpic}[width=0.9\linewidth]{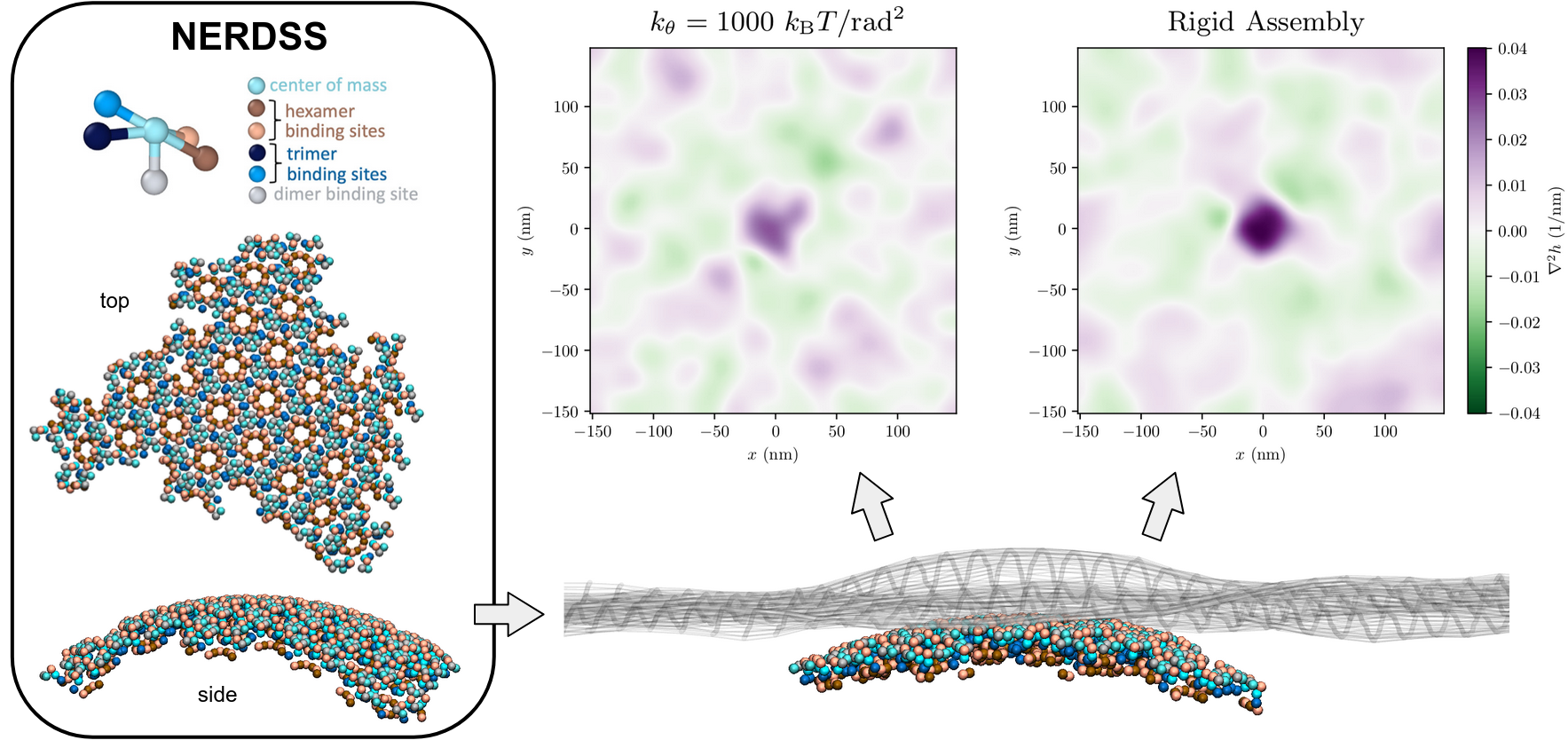}
        \put(0,46){\large \textbf{a} }
        \put(37,12){\large \textbf{b} }
        \put(36,46){\large \textbf{c} }
        \put(67,46){\large \textbf{d} }
    \end{overpic}
    \caption{
        (\textbf{a}) NERDSS CG model of Gag assembly. The Gag monomer and its binding sites are shown first (adapted from Ref. \cite{qian2023temporal}), along with two views of a partially assembled immature lattice generated by NERDSS \cite{qian2023temporal}.
        (\textbf{b}) Snapshot of the NERDSS-assembled lattice simulated as a membrane-coupled flexible assembly using our methods.
        (\textbf{c}) Average local membrane curvature ($K\approx \nabla^2 h$) from $50\,\unit{\micro\second}$ of simulation with all bond parameters set to a numerical value of 1000. The Gag lattice center of mass is centered at the origin.
        (\textbf{d}) The same as (\textbf{c}), but with the Gag lattice treated as a fixed rigid structure. The extent of curvature is roughly double that of the flexible case, with a more uniformly circular membrane footprint. The color scale is the same in both plots.
    }
    \label{fig:gag-curvature-comparison}
\end{figure*}

All of the results presented pertain to a ``bespoke'' flexible clathrin lattice model, where we have manually constructed particular lattice geometries and examined them.
However, as mentioned at the outset, our goal is inter-operability with the NERDSS simulation platform \cite{varga2020nerdss}.
To support this, we have developed a scripted pipeline to convert the structural output of any NERDSS simulation into a ready-to-simulate input for our coupled flexible protein-membrane simulation framework.
The script (\texttt{prepare\_nerdss\_complex.py} in the code repository \cite{code_repo}) parses the supplied NERDSS structure and prompts the user for all required harmonic bond constants, or the user can provide a default value for unknown parameters.
The structure is then ready to be simulated using the methods presented above.
In addition to making the protein assembly flexible, we provide a boolean flag \texttt{rigid\_assembly} for simulating the membrane coupled to the perfectly rigid protein structure as assembled by NERDSS.

As an example, we examine the previously published CG model of HIV immature Gag lattice assembly \cite{qian2023temporal,guo2023structure,ying2026membrane}.
The CG Gag monomer is shown at the top of Fig.~\ref{fig:gag-curvature-comparison}, along with a structure containing 203 Gag monomers that assembled after $\Delta t = 0.83$ seconds in a NERDSS simulation.
Association free energies were constrained by experimental measurements \cite{datta2007interactions}, and rates were chosen within a realistic biological range; see ref. \cite{qian2023temporal} for details of the CG model and NERDSS simulation.
The NERDSS complex file is given as input to our new script along with the name(s) of any CG site which will be bound to the membrane.
For this example, we chose the COM sites on each monomer as the membrane attachment points for simplicity rather than introduce a new CG site.
The equilibrium length of the membrane-Gag bonds was set to $9\,\mathrm{nm}$ based on experimental structural measurements \cite{benjamin2005three,wilk2001organization}.
The structure is oriented to the membrane consistent with the physiology, where the Gag lattice is enveloped by the membrane as it buds out of the cell (the opposite of the clathrin lattice).
The initial positions of the lattice points relative to the membrane surface are found by minimizing the mean-squared perpendicular distance between the membrane-binding sites and the planar membrane at $z=0$.
The structure is then shifted such that all membrane binding sites are on one side of the membrane, choosing the side which results in the membrane binding sites being closer than the average COM site positions (so that it is not bound ``upside-down'').
We carried out two simulations of the Gag lattice-membrane system with the only difference between the two being the rigidity of the lattice: in one case, all bond parameters were set to a numerical value of 1000 ($k_\sigma = 1000 \,\kT/\mathrm{nm}^2$, $k = 1000 \,\kT/\mathrm{rad}^2$ for all angular potentials); in the other case, the lattice is taken to be a perfectly rigid body with no deformation (\texttt{rigid\_assembly=True}).

We simulated both systems for 100 $\unit{\micro\second}$ in a $300\,\mathrm{nm}\times 300\,\mathrm{nm}$ box and analyzed the membrane curvature (in small gradient approximation, $K\approx \nabla^2 h$) for the latter 50 $\unit{\micro\second}$, with the first 50 $\unit{\micro\second}$ discarded for equilibration.
The results are shown in Figs.~\ref{fig:gag-curvature-comparison}c and \ref{fig:gag-curvature-comparison}d.
Giving the Gag lattice perfect rigidity results in more pronounced induced membrane curvature that more closely matches the imposed curvature of the rigid NERDSS lattice.
The resulting membrane curvature also has a more symmetrical pattern, since the scaffolding lattice is a near perfect subsection of a sphere.
In the flexible case the assembly is partially and irregularly flattened, resulting in weaker and less symmetric curvature induction.
The bond parameters here were chosen arbitrarily, however, and do not reflect knowledge of the true underlying Gag lattice rigidity.
Our example serves to demonstrate that this framework allows one to examine the influence of protein lattice rigidity on the induced membrane curvature, both via intrinsic microscopic bond rigidity as well as lattice geometry and topology.

\section{Discussion}

A significant challenge when developing a realistic flexible protein model is the large number of microscopic bond parameters required to fully specify the elasticity of the underlying bond network.
Each degree of freedom in the flexible structure requires a corresponding harmonic stiffness, and these parameters are generally not measurable from experiment as they reflect coarse-grained degrees of freedom not easily mapped to individual protein-protein contacts.
Consequently, designing a quantitatively accurate model requires some means of calibrating the stiffness of the chosen microscopic degrees of freedom to reproduce the emergent and measurable elasticity of the protein assembly, generally at larger length scales.
We have demonstrated two such calibration protocols in this work, namely flat sheet buckling and spherical membrane deformation.
In presenting these, we have highlighted how the specific choice of calibration protocols and accompanying (often unstated) assumptions can make the mapping between micro- and macro-parameters less straightforward than might be expected.
Neither of these protocols is wrong; rather, they demonstrate different aspects of the mechanics of solid-like protein lattices.
Under the stated modeling assumptions, it is then necessary to model different stages of growth with different meso-scale stiffness parameters, rather than assuming a constant lattice stiffness throughout the process.

Our results therefore demonstrate that capturing explicit lattice structure and connectivity can significantly impact the degree of curvature generation of spherical capsid-forming structures.
Care must be taken when attempting to match meso-scale properties of simulation models to experimentally measured rigidities. Unfortunately, the ``correct'' way to calibrate a simulation model may not be one-size-fits-all. Our primary recommendation, which we used as our guiding principle in this work, is to calibrate the simulation model by comparing to the same kind of deformations used in experimental measurements. In our case, this meant examining spherical deformations rather than cylindrical deformations of the protein lattice, as this is the natural clathrin assembly geometry and thus the most commonly measured. An additional unfortunate reality is that this will often not be enough to uniquely determine all of the CG model parameters. In our case, $k_\omega$ is not well-constrained due to it having negligible impact on the measured $\kappa_l$ value and our not possessing separate experimental measurements which quantify the resistance of clathrin lattices to torsional deformations.

A limitation of the modeling framework presented stems from our usage of the small-gradient approximation of the Helfrich functional, Eqn.~(\ref{eqn:small-grad-helfrich}).
The Fourier-space Brownian dynamics formalism used in this work follows from this, and is thus only rigorously valid for membrane shapes with modest gradients.
This approximation accurately captures many biological processes which cause membrane deformation and enables highly efficient simulations, but becomes less reliable as membrane deformations become large.
The greatly simplified equations of motion in FSBD also do not allow for spontaneous curvature, as this would naturally give rise to large membrane height gradients over large length scales.
As such, for applications which require strongly curved membranes, such as complete clathrin cage assembly or immature Gag lattice budding, more robust computational models must be employed, such as Finite Element Method-based approaches\cite{feng2006finite,ying2026membrane,fu2026predicting} or discrete geometry-based approaches\cite{siggel2022trimem,zhu2022mem3dg,brakke1992surface,pezeshkian2024mesoscale} which are designed to more accurately reproduce the fully nonlinear Helfrich model.
While these methods are more complex to implement and have larger computational overhead, the improved accuracy they provide represents an important future direction for this approach.

Another avenue for potential improvement of the presented model is the inclusion of steric repulsion between the membrane and CG binding sites.
As presented, the only interactions between protein subunits and the membrane are the harmonic forces which couple each monomer to the membrane.
For monomers which are small on the scale of the curvature induced by the overall structure (as is the case for our examples), this does not lead to significant issues.
However, curvature induction by individual proteins could entail including additional CG sites representing the bulk protein structure, with corresponding steric repulsions with the membrane.
While we have not included these effects for computational simplicity, there is no barrier to implementing them; they are perfectly compatible with the presented simulation methodology.

When combined with the ioNERDSS python package \cite{ying2026transforming}, which can automatically generate CG models for NERDSS simulations from PDB structures, the methodology presented here completes a computational pipeline spanning experimentally-determined structural data, reaction-diffusion simulation of self-assembly, and the subsequent membrane remodeling driven by the pre-assembled protein complex.
While this current pipeline has de-coupled the reaction-diffusion dynamics from the membrane dynamics, meaning we are not simultaneously capturing recruitment or assembly/disassembly and membrane remodeling, our work here is a critical step towards achieving these more complex, coupled dynamics.
At present, there are two obstacles to fully integrating our framework with NERDSS:
The first is that the propagation of proteins in NERDSS must be modified to account for applied forces, amounting to a drift term in the diffusion process.
The second, more difficult challenge is the modification of reaction rates in NERDSS to account for applied forces in both binding and un-binding.
As it currently stands, the method presented in this work enables the systematic investigation of how molecular-scale binding interactions and lattice mechanics influence membrane deformation during the earliest stages of cellular remodeling processes.
While the current framework does not integrate rigorous (un-)binding kinetics as mentioned, it can still be used to examine dynamic membrane curvature induction by an assembling lattice if the assembly pathway and timescale is prescribed, rather than simulated.
This is the approach taken in recent work by Bouzos, \textit{et al.}\cite{bouzos2026clathrin}, where the present authors modeled membrane curvature induction by an assembling clathrin coat by adding new triskelion monomers to the assembly at fixed time intervals.
Note that, as shown in Fig.~\ref{fig:relaxation-time}, the clathrin curvature relaxation time is on the order of 0.1 to 10 $\unit{\micro\second}$, and the membrane relaxation is on the order of $100\unit{\micro\second}$ (Sec. II.E).
Since clathrin coat assembly happens on the time scale of seconds, the coat–membrane curvature is effectively always at mechanical equilibrium, supporting our quasi-static continuum treatment of growing coats.

The principal advantage of the present framework is its computational efficiency and ease of use, particularly given the mature and well-established FSBD algorithm.
Considerable additional performance improvements could likely be achieved by incorporating the membrane dynamics directly into the HOOMD-blue C++ backend, eliminating much of the current Python-level overhead.
Although the resulting framework remains less general than simulation methods based on the full nonlinear Helfrich theory, it occupies a powerful middle ground by enabling simulations over biologically relevant length and time scales while retaining a direct connection to experimentally-informed CG protein models.

\section*{Supplementary Material}

The supplementary PDF file includes the derivation of Eqn.~(\ref{eqn:buckling_force}), a comparison of buckling in the $\mathbf{\hat{x}}$ vs. $\mathbf{\hat{y}}$ directions, plots of the un-averaged data used to produce Fig.~\ref{fig:buckleresults}, an explanation of our blocking and relaxation timescale error analyses, some additional analytical results related to membrane tension, potential energies of closed clathrin caps with varying rigidity and pucker angle, and a plot of mean membrane profiles for varying $k_\mathrm{mem}$.

We provide two supplementary videos. \texttt{Movie S1} shows approximately the first half of the simulation trajectory for the open cap clathrin structure with $k_\theta=1080\kT/\mathrm{rad}^2$ and membrane rigidity $\kappa_\mathrm{m}=20\kT$. Frames are recorded at 0.5 microsecond intervals. \texttt{Movie S2} shows the early relaxation dynamics of the same system at higher time-resolution, with frames at 10 ns and a total of 10 microseconds shown.

\begin{acknowledgments}
The authors gratefully acknowledge support from NIH grant R35GM161901 to M.E.J., and a postdoctoral award from the Gordon and Betty Moore Foundation to S.L.F.
\end{acknowledgments}

\section*{Author Declarations}
Conflicts of Interest: The authors have no conflicts to disclose.

\section*{Data Availability Statement}
The data that support the findings of this study are available within the article and all code necessary to reproduce the results are freely available at \texttt{github.com/PhysFoley/BrownianProteinFSBDmem}.

\bibliography{refs}
\bibliographystyle{apsrev4-2}

\end{document}


\renewcommand{\theequation}{S\arabic{equation}}
\renewcommand{\thefigure}{S\arabic{figure}}

\maketitle

\section{Buckling Stress-Strain Relation}

\begin{figure}
    \centering
    \includegraphics[width=0.5\linewidth]{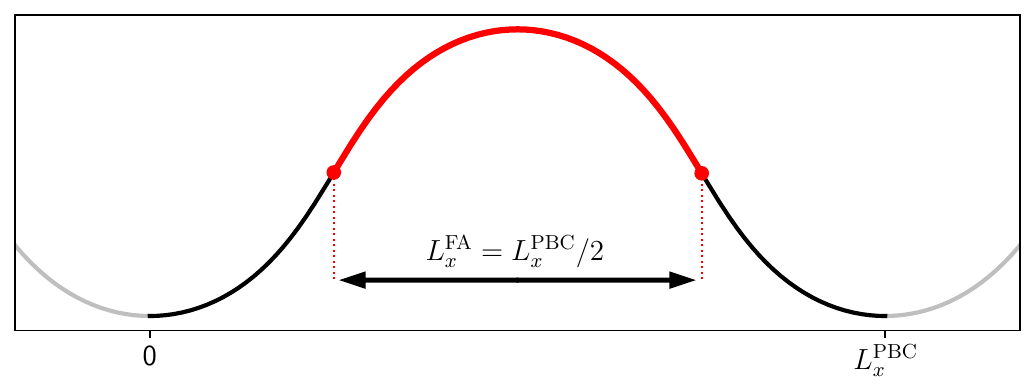}
    \caption{
        Schematic of the two buckling geometries considered. The exact solution of the nonlinear shape equation under PBC (derived in ref. \cite{hu2013determining}) for $\epsilon=0.25$ is shown as the black curve (periodic images shown as faint continued curve). The free angle (FA) BC shape is highlighted in red, with its endpoints (red points) at the inflection points of the PBC shape.
    }
    \label{fig:buckle_bc}
\end{figure}

The authors of ref. \cite{hu2013determining} derive the buckle stress-strain relation $f_x(\epsilon)$ for the case of a continous membrane of length $L$ in the $\hat{x}$-direction which is smoothly periodic in a box with lateral dimensions $L_x,L_y$, with $L_x<L$ (see Fig.~\ref{fig:buckle_bc}).
They do this via a functional variation of the Helfrich energy in angle-arc length parametrization $\psi(s)$ with a Lagrange multiplier $f_x$ to enforce the length $L$ being confined to horizontal length $L_x$. This gives the augmented functional,
\begin{equation}
    \mathcal{E}[\psi] = L_y \int_0^L \mathrm{d}s \left\{ \frac{1}{2}\kappa \dot{\psi}^2 + f_x \left[ \cos\psi - \frac{L_x}{L} \right] \right\},
\end{equation}
whose variation yields
\begin{equation}
    \frac{\delta \mathcal{E}}{L_y} = \left[ \kappa \dot{\psi} \delta \psi \right]_0^L - \int_0^L \mathrm{d}s \left[ \kappa\ddot{\psi} + f_x \sin\psi \right] \delta \psi.
    \label{eqn:variation}
\end{equation}
In their case, the boundary term vanishes since the end points correspond to the same physical location.
From the shape equation $\kappa\ddot{\psi} + f_x \sin\psi=0$ and their periodicity conditions they then derive
\begin{equation}
    f_x^\mathrm{PBC}(\epsilon) = \kappa \left( \frac{2 \pi}{L} \right)^2 \sum_{n=0}^{\infty} b_n \epsilon^n,
\end{equation}
with coefficients $b_n$ given in \cite{hu2013determining}.
However, in the main text, we measure the flexural (buckling) rigidity by buckling a finite non-periodic sheet such that its endpoints are able to pivot freely at fixed position (Fig. 1).
This corresponds to a free-angle (FA) boundary condition at the end points, meaning $\delta \psi$ must be left arbitrary in the boundary term of Eqn.~(\ref{eqn:variation}).
In order for this boundary term to vanish we then must have $\dot{\psi}(0)=\dot{\psi}(L)=0$, that is, the curvature must vanish at the boundaries.
As it happens, this exact geometry already occurs as a subset of the periodic buckle solution.
Fig.~\ref{fig:buckle_bc} highlights the middle portion of the PBC solution, which exactly corresponds to our finite FA buckle.
It spans the region between the two inflection points of the PBC buckle, at which $\dot{\psi}=0$ by definition.
We already know that the geometry between these endpoints satisfies the shape equation, and therefore by the uniqueness of solutions, this subsection is the exact solution for our case.
Since the force in the $x$-direction is constant throughout the shape\cite{hu2013determining}, we know that the same $f_x(\epsilon)$ is the force per unit length required to buckle the \emph{smaller} length $L_x^\mathrm{FA} = L_x^\mathrm{PBC}/2$.
The buckling strain $\epsilon$ is identical for the two shapes, since both $L_x$ and $L$ are halved.
Therefore we have $f_x^\mathrm{PBC}(\epsilon,L) = f_x^{FA}(\epsilon,L/2)$, from which we immediately get the form in Eqn.~(12) of the main text,
\begin{equation}
    f_x^\mathrm{FA}(\epsilon) = \kappa \left( \frac{\pi}{L} \right)^2 \sum_{n=0}^{\infty} b_n \epsilon^n,
\end{equation}
with the same coefficients $b_n$ as in the PBC case.
The net result is a factor 4 decrease in the buckling force.

\section{Buckling along $\hat{x}$ vs. $\hat{y}$}

\begin{center}
    \begin{overpic}[width=0.5\linewidth]{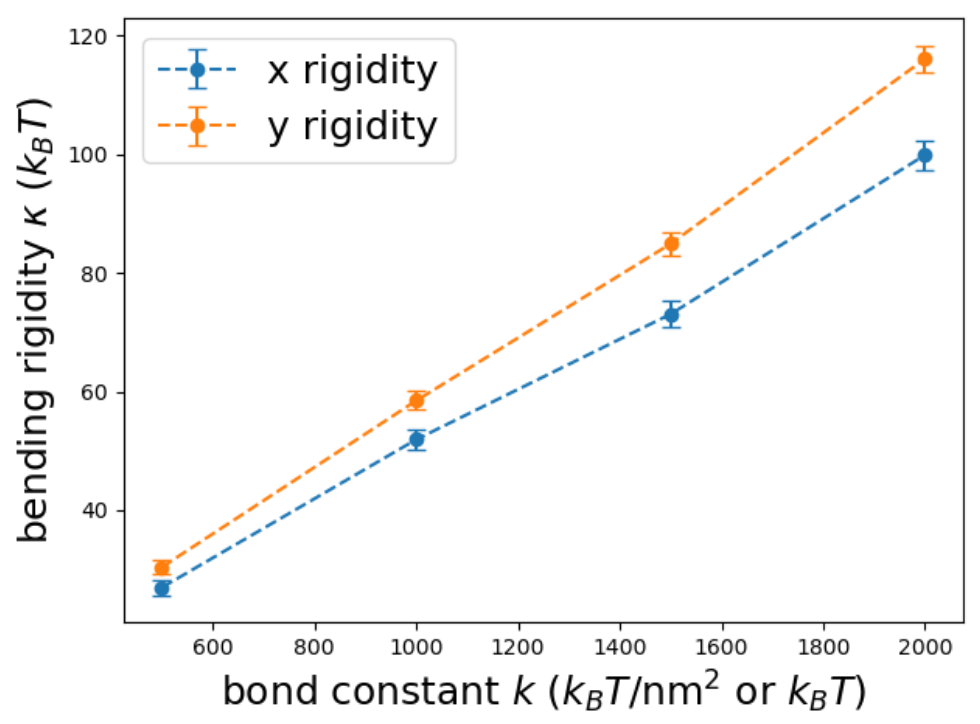}
        \put(70,13){\includegraphics[width=0.13\linewidth]{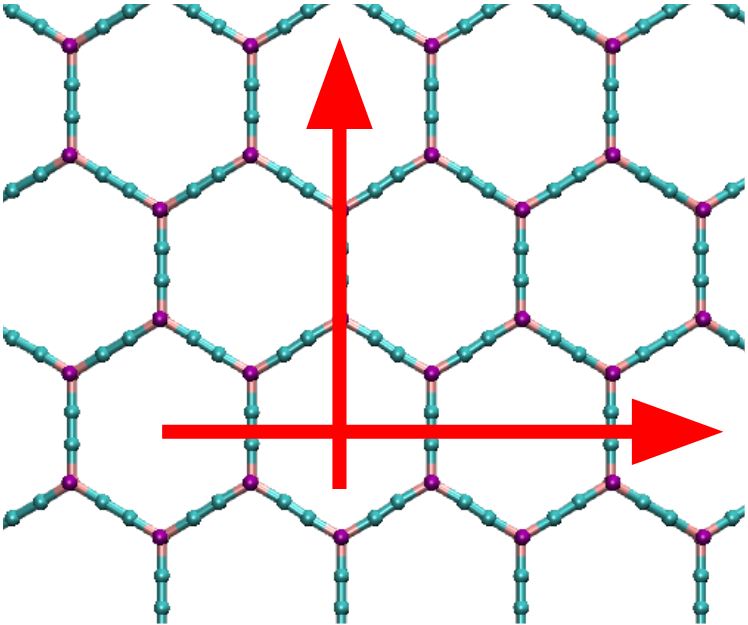}}
    \end{overpic}
    \captionof{figure}{Whether you buckle along $\hat{x}$ or $\hat{y}$ changes the precise value measured for $\kappa_l$, but the results are qualtitatively the same, differing by around 10\% in our tests, as seen here. The lower-right inset illustrates the two non-equivalent bending directions, since the 90-degree rotation between the two does not match the 60-degree symmetry of the lattice.}
    \label{fig:xy_buckling}
\end{center}

\section{Un-Averaged Flexural Rigidity Data}

\begin{center}
    \includegraphics[width=0.7\linewidth]{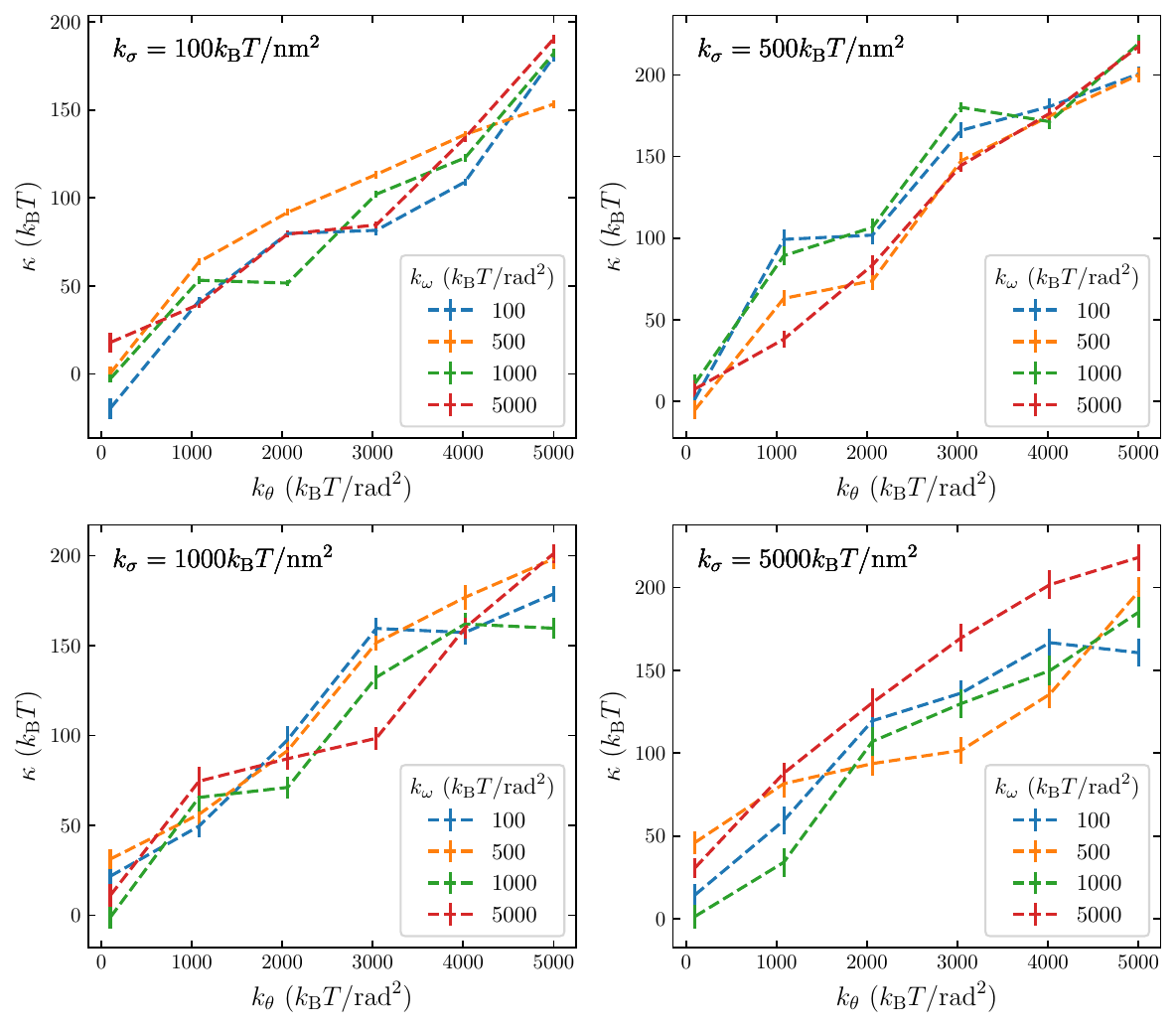}
    \captionof{figure}{Flexural rigidity $\kappa_{\ell,\mathrm{flex}}$ measured from buckling simulations. This figure presents the underlying data used to create Fig. 4 of the main text. Here, each data point comes from a single simulation with error bars determined by the blocking procedure (see next SI section).}
    \label{fig:no-avg-buckle}
\end{center}

\section{Timescales, Correlation, and Blocking Analysis}

\begin{figure}
    \centering
    \includegraphics[width=0.8\linewidth]{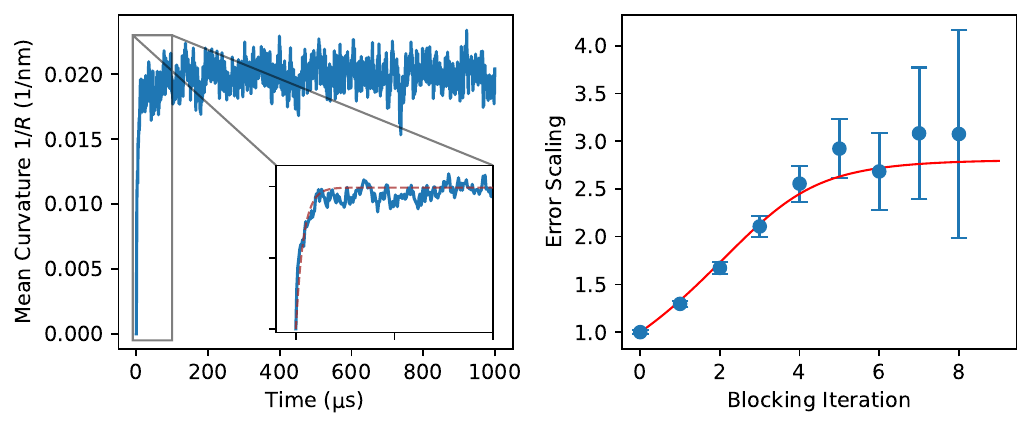}
    \caption{
        Left: Time series of open cap SCC curvature over a complete simulation with $k_\theta = 300 \kT/\mathrm{rad}^2$ and $\kappa_\mathrm{m}=14\kT$. The inset shows the fit of the curvature relaxation to the exponential form of Eqn.~(18) of the main text.
        Right: Scaling of the standard error $(\sigma_k / \sqrt{N_k})/(\sigma/\sqrt{N})$ with blocking iteration for the latter half of the time series shown on the left (blue dots, error bars estimated as explained in Ref. \cite{foley2024quantifying}). The red curve is the fit to Eqn.~(\ref{eqn:blocking-formula}), yielding $T\approx 3.9$, which, given that our simulation time series data was written out at $500\mathrm{ns}$ intervals for this trajectory, gives a correlation time of approximately $2\mu \mathrm{s}$ for the lattice curvature. As can be seen from Fig. 7a of the main text, this closely matches the fitted curvature relaxation timescale $\tau$ for $k_\theta = 300 \kT/\mathrm{rad}^2$, as expected from linear response theory. (Note: for systems with relaxation timescales shorter than $1\mu \mathrm{s}$ seen in Fig. 7 of the main text, the first $10\mu \mathrm{s}$ were re-run with a trajectory output frequency of $5\mathrm{ns}$ to resolve the rapid relaxation dynamics.)
    }
    \label{fig:blocking-fig}
\end{figure}

Throughout this work, error bars for quantities measured from the average of a time series are calculated using a blocking method first elaborated in Ref. \cite{flyvbjerg1989error} and more recently detailed in Ref. \cite{foley2024quantifying}.
Briefly, the method is to take the original time series of length $N$ and average each neighboring pair of values, producing a new time series of length $N/2$.
This process is iterated, producing shorter and shorter time series.
We will refer to the number of times this process has been iterated as $k$, the blocking iteration (the original time series corresponds to $k=0$).
One can show that for an exponentially correlated stationary time series with correlation time $T$ (in units of time series steps), the \textit{naively-calculated} standard error of each of these block-averaged time series ($\sigma_k/\sqrt{N_k}$) grows with blocking iteration $k$ and eventually plateaus at the \textit{true} error of the mean.
The analytical form of this behavior can be derived, yielding\cite{foley2024quantifying},
\begin{equation}
    \frac{\sigma_k / \sqrt{N_k}}{\sigma/\sqrt{N}} = \sqrt{ \frac{1 + c}{1 - c} - \frac{2c}{2^k}\cdot \frac{1 - c^{2^k}}{(1-c)^2}},
    \label{eqn:blocking-formula}
\end{equation}
where $c=e^{-1/T}$ and $\sigma$ is the standard deviation of the original time series data.
Calculating $\sigma_k / \sqrt{N_k}$ for several $k$ iterations and fitting to Eqn.~(\ref{eqn:blocking-formula}) then simulatneously gives the correct error of the mean and an estimate of the correlation time.
An example is shown in Fig.~\ref{fig:blocking-fig}.

\section{Relaxation Timescale Error}

\begin{center}
    \includegraphics[width=0.6\linewidth]{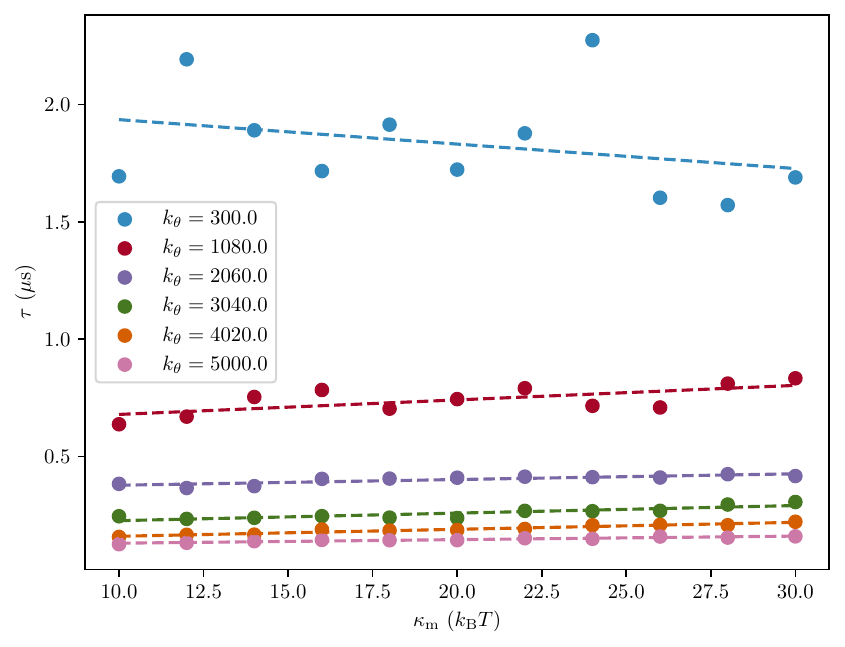}
    \captionof{figure}{Linear fits of $\tau$ vs. $\kappa_\mathrm{m}$ for fixed values of $k_\theta$, for open cap structures at zero membrane tension. $k_\theta$ values are in units of $\kT/\mathrm{rad}^2$. The first line having a negative slope, against the trend of all the others, is likely an artifact of the large scatter of the measurements, in turn due to the large curvature fluctuations for this weak lattice.}
    \label{fig:linear-kappam-fits}
\end{center}

The SCC curvature relaxation timescale $\tau$ in principle depends on both lattice rigidity and membrane rigidity; however, for our simulated systems the influence of $\kappa_\mathrm{m}$ is a small linear trend, as shown explicitly for the open cap simulations at zero tension in Fig.~\ref{fig:linear-kappam-fits}.
We can use this to our advantage to infer a statistical error bar on $\tau$ measured at $\kappa_\mathrm{m}=20\kT$, our mean membrane rigidity.
When fitting $\tau$ via Eqn.~(18) (example shown in Fig.~\ref{fig:blocking-fig}), the uncertainty inferred from the parameter covariance of the least-squares fit (in this case SciPy's \texttt{curve\_fit}) is typically far smaller than the actual run-to-run variation of $\tau$ (these error bars are smaller than the plotting points in Fig.~\ref{fig:linear-kappam-fits}).
This is because the fit treats successive frames as independent and does not account for variation between trajectories.
We therefore treat the $n=11$ simulations at each $k_\theta$, spanning $\kappa_\mathrm{m}=10-30\kT$, as replicates after removing the linear dependence on $\kappa_\mathrm{m}$.
We fit $\tau = a + b\kappa_\mathrm{m}$ by ordinary least-squares and compute the residual variance $s^2 = \frac{1}{n-2} \sum_i (\tau_i - a - b \kappa_{\mathrm{m},i})^2$, where we subtract 2 in the denominator due to our linear fit removing two degrees of freedom.
Because $\kappa_\mathrm{m}=20\kT$ lies at the exact center of the sampled range, the fit value of $\tau$ there equals the sample mean of $\tau$, and its standard error is $s/\sqrt{n}$ \cite{bevington2003data}.
These are the error bars in Fig. 7 of the main text.

\section{Energy of Free Membrane Surrounding Spherical Adhesion}

For the situation of a Helfrich membrane which adheres to and partially wraps a sphere of radius $R$, the author of ref.~\cite{deserno2004elastic} works out the small-gradient energy of the free portion of the membrane to be (in our notation),
\begin{equation}
    \frac{E_\mathrm{free}}{\pi \kappa_\mathrm{m}} = R \sqrt{\frac{\sigma}{\kappa_\mathrm{m}}}\left( \frac{k^3}{1 - k^2} \right) \frac{K_0 \left(kR\sqrt{\sigma/\kappa} \right)}{K_1 \left(kR\sqrt{\sigma/\kappa} \right)},
\end{equation}
where $k=\sqrt{\frac{a}{2\pi R^2} \left( 2 - \frac{a}{2 \pi R^2} \right)}$ and $K_n(x)$ are modified Bessel functions of the second kind \cite{NISTDLMF}.
As this exact form is rather unwieldy, they also work out the expansion of this energy in the limit of small wrapping, which in our case is the limit $a/2\pi R^2 \ll 1$, for which they find,
\begin{equation}
    \frac{E_\mathrm{free}}{\pi \kappa_\mathrm{m}} \approx -\frac{\sigma a^2}{2 \pi R^2} \left[ 2\gamma + \ln\left( \frac{\sigma a}{4 \pi \kappa_\mathrm{m}} \right) \right],
\end{equation}
in which $\gamma\approx 0.5772$ is the Euler-Mascheroni constant.
Since our bending and stretching energies are positive-definite, this approximation can only be valid for parameter values which result in the bracketed expression being negative, thus rendering the overall energy contribution positive.
In any event, under the approximation that the logarithmic term varies little over our parameter regime, we indeed see that the free portion of the membrane contributes an energy proportional to $\sigma a^2 / R^2$ in the appropriate limit.

\section{Optimal Angle for D6 Barrel Clathrin Cage}

\begin{center}
    \includegraphics[width=\linewidth]{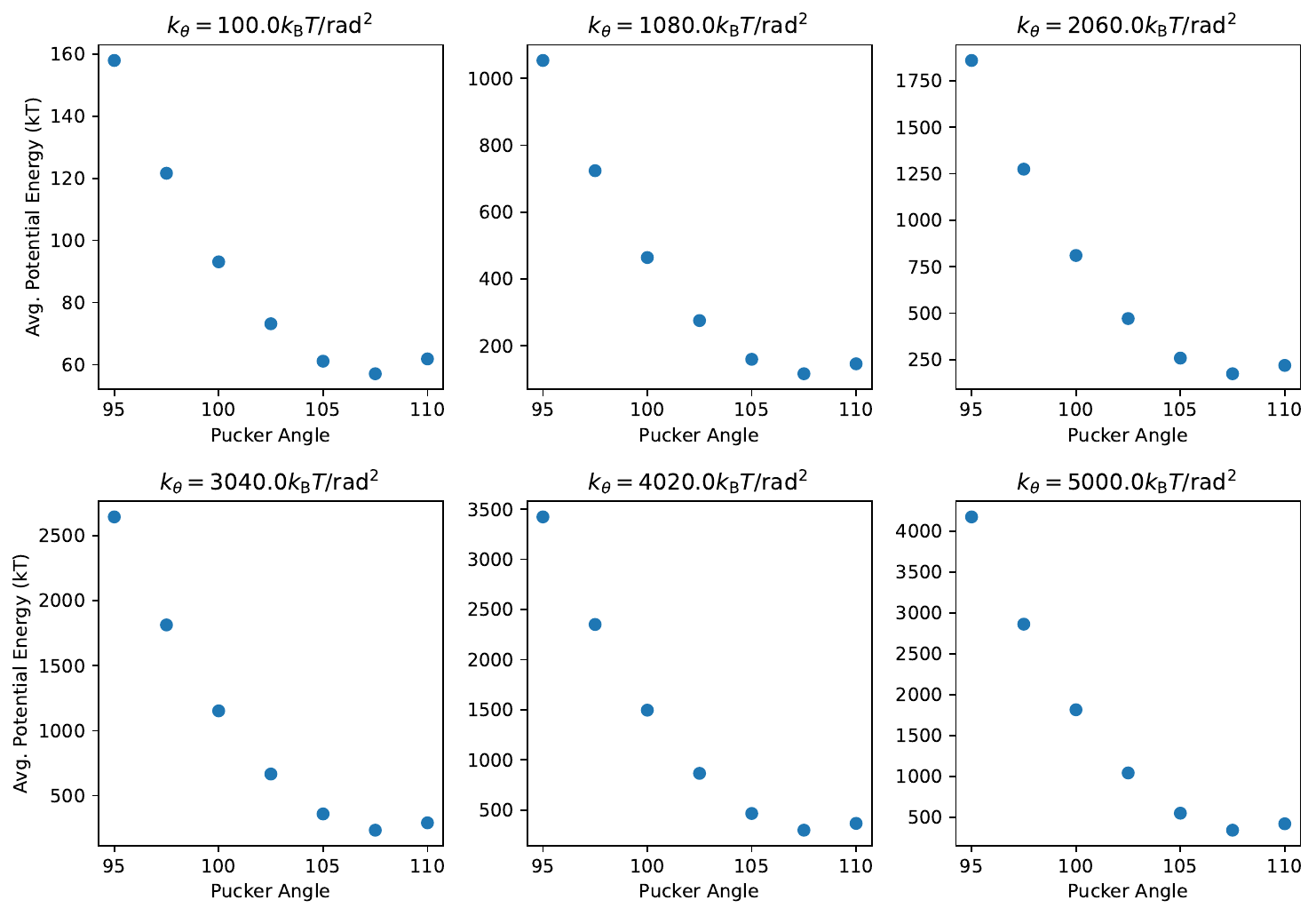}
    \captionof{figure}{Mean potential energy of closed cap clathrin lattices for varying pucker angle and $k_\theta$ as indicated.
    All rigidities have the same energy-minimum pucker angle of $107.5^\circ$.}
    \label{fig:vary_angle}
\end{center}

\section{Membrane Shape with Varying $k_\mathrm{mem}$}

\begin{center}
    \includegraphics[width=0.9\linewidth]{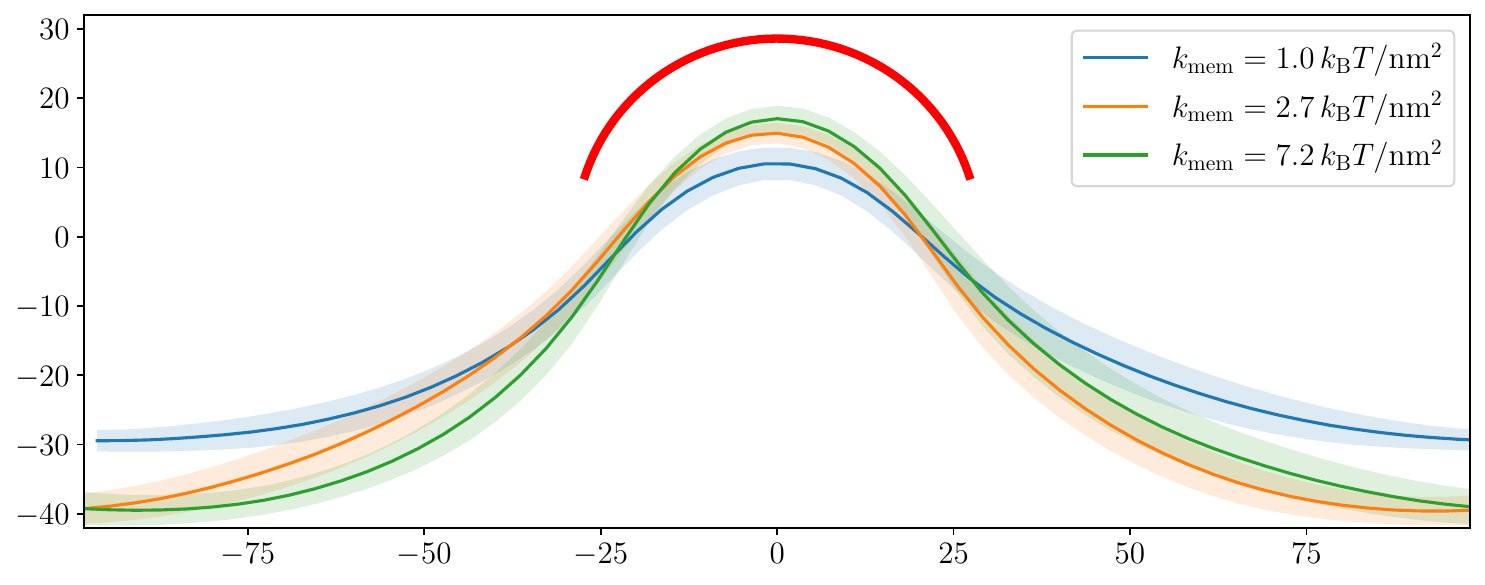}
    \captionof{figure}{For sufficiently small $k_\mathrm{mem}$, the membrane does not strictly conform to the shape of the protein lattice.
    In the above figure, this can be seen for $k_\mathrm{mem}= 1 \,\kT/\mathrm{nm}^2$ (shading shows standard deviation of observed height profile).
    All $k_\mathrm{mem}$ values larger than roughly 3 $\kT/\mathrm{nm}^2$ resulted in shapes that did not differ significantly.}
\end{center}

\printbibliography